\documentclass[aps,nopacs,nokeys,superscriptaddress,11pt]{revtex4}

\usepackage{graphicx,epic,eepic,epsfig,amsmath,latexsym,amssymb,verbatim,revsymb,color}
\usepackage{array}

\usepackage{theorem}
\newtheorem{definition}{Definition}
\newtheorem{proposition}[definition]{Proposition}
\newtheorem{lemma}[definition]{Lemma}

\newtheorem{theorem}[definition]{Theorem}
\newtheorem{corollary}[definition]{Corollary}

\newtheorem{remark}[definition]{Remark}

\def\squareforqed{\hbox{\rlap{$\sqcap$}$\sqcup$}}
\def\qed{\ifmmode\squareforqed\else{\unskip\nobreak\hfil
\penalty50\hskip1em\null\nobreak\hfil\squareforqed
\parfillskip=0pt\finalhyphendemerits=0\endgraf}\fi}
\def\endenv{\ifmmode\;\else{\unskip\nobreak\hfil
\penalty50\hskip1em\null\nobreak\hfil\;
\parfillskip=0pt\finalhyphendemerits=0\endgraf}\fi}
\newlength{\blank}
\newenvironment{proof}[1][{\hspace{-\blank}}]{{\noindent\textbf{Proof~{#1}.\ }}}{\hfill\qed\vskip 0.5\baselineskip}

\DeclareRobustCommand\idop{%
	\leavevmode\hbox{\small1\normalsize\kern-.33em1}}

\DeclareRobustCommand\scriptidop{\leavevmode\hbox{\fontsize{7}{8}\selectfont 1\scriptsize\kern-.33em1}}

\mathchardef\ordinarycolon\mathcode`\:
\mathcode`\:=\string"8000
\def\vcentcolon{\mathrel{\mathop\ordinarycolon}}
\begingroup \catcode`\:=\active
  \lowercase{\endgroup
  \let :\vcentcolon
  }

\newcommand{\nc}{\newcommand}
\nc{\rnc}{\renewcommand}
\nc{\beq}{\begin{equation}}
\nc{\eeq}{{\end{equation}}}
\nc{\beqa}{\begin{eqnarray}}
\nc{\eeqa}{\end{eqnarray}}
\nc{\lbar}[1]{\overline{#1}}
\nc{\bra}[1]{\langle#1|}
\nc{\ket}[1]{|#1\rangle}
\nc{\ketbra}[2]{|#1\rangle\!\langle#2|}
\nc{\braket}[2]{\langle#1|#2\rangle}
\nc{\inp}[2]{\langle#1|#2\rangle}
\nc{\proj}[1]{| #1\rangle\!\langle #1 |}
\nc{\avg}[1]{\langle#1\rangle}
\nc{\Rank}{\operatorname{Rank}}
\nc{\smfrac}[2]{\mbox{$\frac{#1}{#2}$}}
\nc{\tr}{\operatorname{Tr}}
\nc{\ox}{\otimes}
\nc{\Ox}{\bigotimes}
\nc{\dg}{\dagger}
\nc{\dn}{\downarrow}
\nc{\cA}{{\cal A}}
\nc{\cB}{{\cal B}}
\nc{\cC}{{\cal C}}
\nc{\cD}{{\cal D}}
\nc{\cE}{{\cal E}}
\nc{\cF}{{\cal F}}
\nc{\cG}{{\cal G}}
\nc{\cH}{{\cal H}}
\nc{\cI}{{\cal I}}
\nc{\cJ}{{\cal J}}
\nc{\cK}{{\cal K}}
\nc{\cL}{{\cal L}}
\nc{\cM}{{\cal M}}
\nc{\cN}{{\cal N}}
\nc{\cO}{{\cal O}}
\nc{\cP}{{\cal P}}
\nc{\cR}{{\cal R}}
\nc{\cS}{{\cal S}}
\nc{\cT}{{\cal T}}
\nc{\cX}{{\cal X}}
\nc{\cZ}{{\cal Z}}
\nc{\csupp}{{\operatorname{csupp}}}
\nc{\qsupp}{{\operatorname{qsupp}}}
\nc{\var}{\operatorname{var}}
\nc{\rar}{\rightarrow}
\nc{\lrar}{\longrightarrow}
\nc{\polylog}{\operatorname{polylog}}
\nc{\1}{{\openone}}
\nc{\sign}{{\operatorname{sign}}}

\def\p{\pi}

\def\s{\sigma}

\nc{\RR}{{{\mathbb R}}}
\nc{\CC}{{{\mathbb C}}}
\nc{\FF}{{{\mathbb F}}}
\nc{\NN}{{{\mathbb N}}}
\nc{\ZZ}{{{\mathbb Z}}}
\nc{\PP}{{{\mathbb P}}}
\nc{\QQ}{{{\mathbb Q}}}
\nc{\UU}{{{\mathbb U}}}
\nc{\EE}{{{\mathbb E}}}
\nc{\id}{{\operatorname{id}}}

\nc{\olambda}{{\overline{\lambda}}}
\nc{\ulambda}{{\underline{\lambda}}}

\nc{\be}{\begin{equation}}
\nc{\ee}{{\end{equation}}}
\nc{\bea}{\begin{eqnarray}}
\nc{\eea}{\end{eqnarray}}
\nc{\<}{\langle}
\rnc{\>}{\rangle}
\nc{\Hom}[2]{\mbox{Hom}(\CC^{#1},\CC^{#2})}
\nc{\rU}{\mbox{U}}

\nc{\ob}[1]{#1}

\def\Pvphi{\ketbra{\varphi}{\varphi}}
\def\Ppsi{\ketbra{\psi}{\psi}}

\newcommand{\inlinediag}[1]{\parbox{1in}{\includegraphics[width=1in]{#1}}}
\newcommand{\inlinediagm}[2]{\parbox{#2}{\includegraphics[width=#2]{#1}}}

\newcommand{\assign}{\ensuremath{\kern.5ex\raisebox{.1ex}{\mbox{\rm:}}\kern -.3em =}}

\newcommand{\pt}{^{\Gamma}} 
\def\hc{^{\dagger}} 

\def\L{\left}
\def\R{\right}

\begin{document}

\title{Distinguishability of quantum states under restricted families of measurements
\protect\\ with an application to quantum data hiding}

\author{William Matthews}
\affiliation{Department of Mathematics, University of Bristol, Bristol BS8 1TW, U.K.}
\email{william.matthews@bris.ac.uk}

\author{Stephanie Wehner}
\affiliation{Institute for Quantum Information, Caltech, Pasadena CA 91125, USA }
\email{wehner@caltech.edu}

\author{Andreas Winter}
\affiliation{Department of Mathematics, University of Bristol, Bristol BS8 1TW, U.K.}
\affiliation{Centre for Quantum Technologies, National University of Singapore,
 2 Science Drive 3, Singapore 117542}
\email{a.j.winter@bris.ac.uk}

\date{10 October 2008}

\begin{abstract}
The starting point of the present investigation is the well-known result by 
Helstrom, identifying the best achievable bias in distinguishing two
quantum states under all measurements as (half) the trace norm of
their difference.
We turn this around, noticing that every sufficiently rich set $\mathbf{M}$ of
measurements on a fixed quantum system defines a statistical
norm $\|\cdot\|_{\mathbf{M}}$ on the states of that system, via the optimal
bias achievable when restricted to $\mathbf{M}$.
These norms are all upper bounded by the usual trace norm, and in finite
dimensional Hilbert spaces they are all equivalent to the
trace norm in the sense that there exist ``constants of domination'' $\lambda$ and $\mu$,
such that
\[
  \lambda \|\cdot\|_1 \leq \|\cdot\|_{\mathbf{M}} \leq \mu \|\cdot\|_1
\]
which are optimal in the sense that there exist states such that the bounds are tight.
In other words, if we rate the performance of a set of measurements in distinguishing a
given pair of states (of equal prior probability) as the ratio of the largest bias that can 
be obtained by such measurements 
to the best bias achievable when allowing \emph{all} measurements, then $\lambda$ and $\mu$ determine the worst and best case performance respectively for
any pair of states.
Here we set ourselves the task of computing, or at least bounding such constants
for various sets of measurements $\mathbf{M}$.

Specifically, we look at the case that $\mathbf{M}$
consists only of a single measurement, namely the uniformly random POVM, $2$-designs, and $4$-designs
where we find asymptotically
tight bounds for $\lambda$ and $\mu$.
Furthermore, we analyse the multipartite 
setting where the
set of measurements consists of all POVMs implementable by
local operations and classical communication (among other, related classes). 

In the case of two parties, we
show that the lower domination constant $\lambda$ is the same
as that of a tensor product of local uniformly random
POVMs up to a constant. This answers in the affirmative an open question about the
(near-)optimality of bipartite data hiding: The bias that can be achieved by
LOCC in discriminating two orthogonal states of a $d \times d$ bipartite system is $\Omega(1/d)$,
which is known to be tight.
Finally, we use our analysis to derive certainty relations 
(in the sense of Sanchez-Ruiz) for any such measurements and to lower bound
the locally accessible information for bipartite systems.
\end{abstract}

\maketitle

\section*{INTRODUCTION}
Quantum measurements, as described by positive operator valued
measures (POVMs) $(M_k) \assign (M_k)_{k=1}^n$ on some Hilbert space ${\cal H}$,
can be viewed as completely positive 
maps from
the set of density operators to probability vectors
\[
  {\cal M}: \xi \longmapsto \sum_{k=1}^n \proj{k} \tr(\xi M_k),
\]
where we represent the probability vector as an
operator, diagonal in the ``label'' basis $\{\ket{k}\}$ of the $n$-dimensional
space $\CC^n$.
As such, they are non-increasing for the statistical distance
given by the trace norm $\| X \|_1 = \tr |X| = \tr \sqrt{X^\dagger X}$:
since ${\cal M}$ is a completely positive
and trace preserving (CPTP) map, we have that
for any operator $X$ 
\begin{equation}
  \label{eq:contraction}
  \| {\cal M}(X) \|_1 \leq \| X \|_1.
\end{equation}
If $X \geq 0$, then equality holds as both the left, and the right hand side
equal $\tr (X)$. However, for traceless Hermitian operators $X$, which are all
-- up to a scalar factor -- of
the form $\rho-\sigma$ with states $\rho$ and $\sigma$ (i.e.~positive semidefinite
operators of trace $1$), the inequality
\begin{equation}
  \label{eq:contraction-0}
  \| {\cal M}(\rho-\sigma) \|_1 \leq \| \rho-\sigma \|_1
\end{equation}
is typically strict. On the other hand, there always exists a measurement that saturates
this inequality -- for instance the spectral measure of $\rho-\sigma$. In fact,
the two-outcome projective measurement $(M_0,\1-M_0)$ with $M_0$ a projector onto the positive
eigenspace of $\rho - \sigma$ makes eq.~(\ref{eq:contraction-0}) an equality.
This is essentially the content of Helstrom's Theorem: consider a situation
where we want to discriminate between two (a priori equiprobable)
states $\rho$ and $\sigma$. Then, for a given measurement POVM $(M_k)$,
we base our decision on the probability vectors $\vec{p} = \bigl( \tr(\rho M_k) \bigr)$ and
$\vec{q} = \bigl( \tr(\sigma M_k) \bigr)$.
The optimal decision rule is easily seen to be the \emph{maximum
likelihood rule}: observing $k$, we decide on $\rho$ if $p_k > q_k$,
otherwise on $\sigma$. This gives us error probability
\[
  P_E = \frac{1}{2}-\frac{1}{4}\sum_k |p_k-q_k|
      = \frac{1}{2}-\frac{1}{4} \| \vec{p}-\vec{q} \|_1
      =  \frac{1}{2}-\frac{1}{4} \| {\cal M}(\rho-\sigma) \|_1,
\]
where for the probability vectors we refer to the $\ell^1$-norm
and for the image of ${\cal M}$ to the trace norm, since they coincide
for diagonal matrices.
The quantity
\[
  \beta\bigl( (M_k); \rho-\sigma \bigr)
  := \frac{1}{2} \| \vec{p}-\vec{q} \|_1 = \frac{1}{2} \| {\cal M}(\rho-\sigma) \|_1,
\]
ranging between $0$ (for identical probability vectors) and $1$
(for orthogonal probability vectors)
is known as the \emph{bias} of the POVM on the state pair $(\rho,\sigma)$.
Helstrom's Theorem is the statement 
that
the minimum error probability over \emph{all} POVMs is
$\frac{1}{2}-\frac{1}{4} \| \rho-\sigma \|_1$, corresponding to the
maximum bias being $\frac{1}{2} \| \rho-\sigma \|_1$.

Thus we are motivated to look in generality at the situation
where we are restricted in our choice of measurement. Formally,
let $\mathbf{M}$ be a set of POVMs. For convenience, take the elements of $\mathbf{M}$ to be discrete POVMs, since 
we will show below
that we lose nothing by restricting to two-outcome
POVMs. The optimal bias achievable by $\mathbf{M}$ 
for $\xi = \rho-\sigma$ is given by $\frac{1}{2} \| \xi \|_{\mathbf{M}}$ where
the norm is defined by
\begin{equation}
  \label{eq:M-norm}
  \| \xi \|_{\mathbf{M}} := \sup_{(M_k) \in \mathbf{M}} \| {\cal M}(\xi) \|_1.
\end{equation}

In Section~\ref{sec:defi},
we start by recording a few simple implications and properties of this definition.
First of all, we will see that it really is a norm.
We make a connection to general norms in vector spaces, showing in particular that
any norm on trace class operators can be interpreted as a norm of the above
type. We then turn to a number of particular examples, highlighting especially
the problem of determining the constants of domination of $\| \cdot \|_{\mathbf{M}}$
with respect to $\| \cdot \|_1$.
In Section~\ref{sec:single}, we investigate the particular case where $\mathbf{M}$ 
consists of only one (necessarily informationally complete) POVM, finding
the best constants of domination. These constants are attained for the isotropic
(unitary invariant) POVM. We also show how to analyse the situation for
POVMs originating from $2$- and $4$-designs.
In Section~\ref{sec:local-POVM}, we look at the situation that the system under
consideration is bi- or multipartite, and that the POVMs are restricted
to classes respecting the partition: local measurements,
with or without classical communication between the parties,
and extensions of this class. The existence of data hiding~\cite{data-hiding1,data-hiding2,rand} states
yields bounds on the constants of domination in one direction.
Most notably, we show here that in the bipartite case, 
these bounds are optimal up to a constant
factor by analysing the tensor product of two isotropic local POVMs:
it turns out that the resulting measurement attains almost the same bias. 
Hence, the hiding states of~\cite{data-hiding1} are already (near) optimal in the sense
that we cannot hope to construct states which are less well distinguishable under LOCC operations.
In Section~\ref{sec:certainty}, we make a connection to
Sanchez-Ruiz' ``certainty relations'' for mutually unbiased 
bases~\cite{Sanchez-Ruiz:certainty}, which we show holds more generally
for any 2-design POVM, and -- even in a stronger form -- for 4-designs.
We also show how our results for bipartite systems imply a universal
lower bound on the information accessible by LOCC from any pure state
ensemble.
Several appendices contain the proofs of more technical results in the main text.

\section{First observations on norms and dual norms}
\label{sec:defi}

Before turning to the essential observations that we will need later on, we first
explain some basic concepts.
At the heart of the Helstrom-Holevo Theorem on optimal state discrimination
lies the duality between the operator norm $\| \cdot \|$ and the trace norm
$\| \cdot \|_1$:
For operators $\alpha$, $A$ on a Hilbert space ${\cal H}$, these are dual to each other with
\begin{align*}
  \| \alpha \|_1 &= \sup_{\| B \| \leq 1} | \tr (\alpha^\dagger B) |, \\
  \| A \|        &= \sup_{\| \beta \|_1 \leq 1} | \tr(\beta^\dagger A) |.
\end{align*}
In finite dimension, which we shall assume throughout this paper, the
suprema are easily seen to be maxima.
The duality persists when we restrict to Hermitian
(self-adjoint) operators $\alpha = \alpha^\dagger$, $A = A^\dagger$:
\begin{align*}
  \| \alpha \|_1 &= \max_{B=B^\dagger,\ \| B \| \leq 1}               \tr (\alpha B), \\
  \| A \|        &= \max_{\beta=\beta^\dagger,\ \| \beta \|_1 \leq 1} \tr (\beta A).
\end{align*}
These equations are direct consequences of the singular value decomposition
in the general, and of the spectral theorem in the Hermitian case.

The role of the Hilbert-Schmidt inner product, which makes the
real vector space of Hermitian operators, ${\cal B}^{\rm sa}({\cal H})$,
a Euclidean space, becomes more evident in geometrical language by
saying that the unit balls
\begin{align*}
  B_1\bigl( \| \cdot \|_1 \bigr) &= \bigl\{ \alpha=\alpha^\dagger : \| \alpha \|_1 \leq 1 \bigr\}, \\
  B_1\bigl( \| \cdot \| \bigr)   &= \bigl\{ A=A^\dagger : \| A \| \leq 1 \bigr\},
\end{align*}
are \emph{polar} to each other. To explain this notion, note that
the unit ball of any norm $N$ on a finite dimensional real vector space,
\[
  K := B_1(N) = \{ x : N(x) \leq 1 \},
\]
is a topologically closed, convex and symmetric set (i.e.~$K=-K$),
containing the origin $0$ in its interior. Any such body $K$ conversely
determines a norm
\[
  \| x \|_{\check{K}} = \inf \left\{ \frac{1}{t} : t>0 \text{ and } tx \in K \right\},
\]
and it is immediately verified that $K = B_1(\|\cdot\|_{\check{K}})$ and
$N = \| \cdot \|_{\check{K}}$.
That is, norms and convex closed symmetric bodies
of full dimension are equivalent descriptions. Now, the polar of
$K$ in a Euclidean vector space with inner product $\langle \cdot , \cdot \rangle$
is defined to be
\[
  \check{K} := \{ y : \forall x\in K\ \langle x, y \rangle \leq 1 \}.
\]
It is easy to verify that if $K$ is symmetric, convex and
closed, and contains the origin in its interior, then $\check{K}$
has the same properties, and $\check{\check{K}} = K$.

By the above discussion, $K$ is the unit ball of $\|\cdot\|_{\check{K}}$, $\check{K}$
is the unit ball of $\|\cdot\|_K$ and one
has the important, but elementary, formulas
\begin{align*}
  \| y \|_K           &= \max_{x\in K} \langle x,y \rangle, \\
  \| x \|_{\check{K}} &= \max_{y\in \check{K}} \langle x,y \rangle.
\end{align*}
which are the abstract versions of the equations above.

We are now ready to make a series of observations. First, we need to show
that eq.~(\ref{eq:M-norm}) really does constitute a norm for trace class operators, i.e.
for operators with a finite, well-defined, trace. We thereby call a set of POVMs $\mathbf{M}$ \emph{separating}, or ``\emph{informationally complete}''
if for any nonzero operator $\xi\neq 0$, there exists a POVM $(M_k)\in\mathbf{M}$
  and an index $k_0$ such that $\tr (\xi M_{k_0}) \neq 0$.
\begin{lemma}
  \label{lemma:M-norm}
  Eq.~(\ref{eq:M-norm}) above defines a norm on the set of trace class operators
if and only if the set of POVMs is separating.
  Furthermore, $\| \cdot \|_{\mathbf{M}} \leq \| \cdot \|_1$.
\end{lemma}
\begin{proof}
An immediate consequence of the fact that $\| \cdot \|_1$ is a norm, and 
eq.~(\ref{eq:contraction-0}).
\end{proof}

We now show that we can restrict ourselves to POVMs with 2 outcomes. Intuitively, since
we decide between two options (e.g. $\rho$ and $\sigma$ above), we can group the
outcomes of each POVM in two. It is then not difficult to verify that

\begin{lemma}

 \label{lemma:2-outcome}

 For any separating set $\mathbf{M}$ of POVMs $(M_k)_{k=1}^n$ the set
of $2$-outcome POVMs,

$$
\mathbf{M}_2 := \bigl\{ (M,\1-M) : \exists (M_k)_{k=1}^n \in\mathbf{M},\
\exists I \subseteq [n] \quad M=\sum_{k\in I} M_k \bigr\},
$$
satisfies $\| \cdot \|_{\mathbf{M}} = \| \cdot \|_{\mathbf{M}_2}$.
Furthermore, the set
$$
\mathbb{M} := \overline{ \operatorname{conv}\, \{ 2M-\1 : (M,\1-M)
\in \mathbf{M}_2 \} }
$$
is a (closed) symmetric convex body, contained in the operator interval
$[-\1;\1] = \{ X : -\1 \leq X \leq \1 \}$ and containing
$\pm\1$, and is of full dimension, such that
\[
\| \xi \|_{\mathbf{M}} = \max_{M \in \mathbb{M}} |\tr (\xi M)| =: \|
\xi \|_{\mathbb{M}}.
\]
\end{lemma}
\begin{proof}
       For a given $\xi = \rho - \sigma$ and POVM $(M_k)_{k=1}^n \in \mathbf{M}$, the
       bias achieved is simply $\sum_{k=1}^n | \tr \L( M_k \xi \R) |$.

       The same bias is achieved by the $2$-outcome POVM $\L( M_+, M_- \R)
\in \mathbf{M}_2$, where
       \begin{align*}
               M_+ = \sum_{k \in P} M_k, P = \{k \in [n] : \tr(M_k \xi) \geq 0\},\\
               M_- = \sum_{k \in N} M_k = \openone - M_+, N = \{k \in [n] : \tr(M_k
\xi) < 0\}
       \end{align*}
       and clearly no other grouping of the elements of this POVM can result
in a larger bias.

\end{proof}

Note that $\mathbb{M}$ has a non-empty interior (and then contains
  the origin in its interior) if and only if the collection
  $\mathbf{M}$ is informationally complete, which the case if and only
  if $\mathbf{M}_2$ is informationally complete. Mathematically
  the information-completeness is expressed by $\mathbb{M}$,
  spanning the whole operator space. Furthermore, note that from our discussion above we have that

\begin{remark}
  \label{rem:duality}
  The symmetric convex body $\mathbb{M}$ defines two norms, one on the observables
  and effects, the other on the trace class operators, via
  \begin{align}
    \| M \|_{\check{\mathbb{M}}}
                           &= \inf\left\{ \frac{1}{t} : t>0 \text{ and }tM \in \mathbb{M} \right\}, \\
    \| \xi \|_{\mathbb{M}} &= \max_{M \in \mathbb{M}} \tr (\xi M).
  \end{align}
  The first has 
  exactly $\mathbb{M}$ as its unit ball, the second has as its unit ball the
  \emph{polar} of $\mathbb{M}$, i.e.
  \[
    \check{\mathbb{M}} = \bigl\{ \xi : \forall M\in\mathbb{M}\ \tr(\xi M) \leq 1 \bigr\}.
  \]
  The norm $\| \cdot \|_{\mathbb{M}} (= \| \cdot \|_{\mathbf{M}})$
  is dual to $\| \cdot \|_{\check{\mathbb{M}}}$:
  \begin{align*}
    \| \xi \|_{\mathbb{M}} &= \max\bigl\{ \tr(\xi M) : \| M \|_{\check{\mathbb{M}}} \leq 1 \bigr\}, \\
    \| M \|_{\check{\mathbb{M}}}
                           &= \max\bigl\{ \tr(\xi M) : \| \xi \|_{\mathbb{M}} \leq 1 \bigr\}.
  \end{align*}
\end{remark}

Putting everything together, we can now see that
\begin{theorem}
  \label{thm:norm-classification}
  The norms $\| \cdot \|_{\mathbf{M}}$ associated to sets of
  POVMs are in one-to-one correspondence with full-dimensional
  symmetric closed 
  convex bodies $\pm\1 \in \mathbb{M} \subseteq [-\1;\1]$.

  As a consequence, \emph{any} norm $\|\!| \cdot \|\!| \leq \| \cdot \|_1$
  can be written as $\|\!| \cdot \|\!| = \| \cdot \|_{\mathbf{M}}$ for
  some set of POVMs.
\end{theorem}
\begin{proof}
First, starting with a set of POVMs $\mathbf{M}$ defining norms $\|\cdot\|_{\mathbf{M}}$,
Lemma~\ref{lemma:2-outcome}
describes how to construct $\mathbb{M}$, such that
$\| \cdot \|_{\mathbf{M}} = \| \cdot \|_{{\mathbb{M}}}$.

Conversely, starting with a full-dimensional symmetric closed convex body $\mathbb{M} \subseteq [-\1;\1]$, we can construct
a set of POVMs
 $\mathbf{M} = \{ (M,\1-M) : M \in \mathbb{M} \text{ and } M\geq 0 \}$ for which
$\| \cdot \|_{\mathbf{M}} = \| \cdot \|_{{\mathbb{M}}}$.
\end{proof}

We formalise the connection with the state
discrimination problem in the following theorem.
\begin{theorem}
  \label{thm:M-norm:bias}
  Let $\mathbf{M}$ be a set of POVMs on a given Hilbert space, and let
  $\mathbf{M}_2$ and $\mathbb{M}$ be defined as above. For any two states
  $\rho$ and $\sigma$, consider the minimum error probability 
  $P_E^{\mathbf{M}}$ of discriminating between these (a priori equiprobable states).
  Then,
  \[
    P_E^{\mathbf{M}} = \inf_{(M,\scriptidop-M) \in \mathbf{M}_2}
                                \frac{1}{2} - \frac{1}{2} | \tr ((\rho-\sigma)M) |
                     = \frac{1}{2} - \frac{1}{4} \| \rho-\sigma \|_{{\mathbb{M}}}.
  \]
  That is, $\frac{1}{2}\| \rho-\sigma \|_{\mathbb{M}}$ is the bias achievable
  in discriminating $\rho$ from $\sigma$ when only measurements
  in $\mathbf{M}$ are allowed.
  \qed
\end{theorem}

In finite dimension, which is the case we stick to in this paper,
the operators also form a finite-dimensional space, and all these norms are
``equivalent'' in the sense that there are $\lambda',\mu' >0$ such that
\begin{equation}
  \label{eq:norm-domination}
  \lambda' \| \cdot \|_1 \leq \| \cdot \|_{{\mathbb{M}}}
                                           \leq \mu' \| \cdot \|_1.
\end{equation}
By using the above correspondences and dualities, we see that this
is equivalent to
\begin{equation}
  \label{eq:unitball-domination}
  \lambda' [-\1;\1] \subseteq \mathbb{M} \subseteq \mu' [-\1;\1].
\end{equation}

We will use $\lambda_1(\mathbb{M})$ ($\mu_1(\mathbb{M})$) to denote
the largest $\lambda'$ (smallest $\mu'$) in these equations.
The numbers $\lambda_1$ and $\mu_1$ are called
the \emph{constants of domination} of the norm
$\| \cdot \|_{{\mathbb{M}}}$ (with respect to $\| \cdot \|_1$).
In the following, our goal is to bound these constants
of domination for various interesting classes of POVMs. 
These constants are especially interesting, since we know from Theorem~\ref{thm:M-norm:bias} that they allow us
to bound the bias that we can achieve when trying to distinguish two states $\rho$ and $\sigma$
with a restricted set of measurements.

Note that 
$\mu_1(\mathbb{M})$ is trivially $1$ since for $\rho \geq 0$,
$\| \rho \|_{\mathbb{M}} = \| \rho \|_1 = \tr (\rho)$. Thus, we are
motivated to restrict to \emph{traceless} operators in eq.~(\ref{eq:norm-domination}). This is also 
the setting for which bounds on the constants of domination give us
a bound on the bias of distinguishing two a priori equiprobable states $\rho$ and $\sigma$.
Let $\lambda(\mathbb{M})$ and $\mu(\mathbb{M})$ be the largest
and smallest numbers $\lambda'$ and $\mu'$, respectively, such that
\begin{equation}
  \label{eq:norm-domination-0}
  \forall \xi \text{ with }\tr(\xi)=0\quad
  \lambda \| \xi \|_1 \leq \| \xi \|_{{\mathbb{M}}} \leq \mu \| \xi \|_1.
\end{equation}
Equivalently, in the dual picture we have to go to the quotient modulo
multiples of the identity, $\RR\1$:
\begin{equation}
  \label{eq:unitball-domination-0}
  \lambda [-\1;\1]/_{\RR\scriptidop} \subseteq \mathbb{M}/_{\RR\scriptidop} \subseteq \mu [-\1;\1]/_{\RR\scriptidop}.
\end{equation}
The following lemma characterizes $\lambda_1$ ($\mu_1$) and $\lambda$ ($\mu$), and their respective relations.

\begin{lemma}
  \label{rem:optimisation}
  For a set $\mathbf{M}$ of POVMs with associated convex body $\mathbb{M}$,
  the constants of domination can be expressed as the solutions of the
  following optimisation problems:
  \begin{align*}
    \frac{1}{2}\lambda(\mathbb{M}) \leq
    \lambda_1(\mathbb{M}) =    \inf_{\|\xi\|_1=1} \sup_{(M_k)\in\mathbf{M}} \| {\cal M}(\xi) \|_1
                         &\leq \inf_{{\|\xi\|_1=1 \atop \tr(\xi)=0}} 
                                                  \sup_{(M_k)\in\mathbf{M}} \| {\cal M}(\xi) \|_1 
                          =    \lambda(\mathbb{M}),  \\
    1 =
    \mu_1(\mathbb{M})     = \sup_{\|\xi\|_1=1} \sup_{(M_k)\in\mathbf{M}} \| {\cal M}(\xi) \|_1
                         &\geq \sup_{{\|\xi\|_1=1 \atop \tr(\xi)=0}} 
                                                  \sup_{(M_k)\in\mathbf{M}} \| {\cal M}(\xi) \|_1 
                          =    \mu(\mathbb{M}).
  \end{align*}
  Here, for the purpose of $\lambda$ and $\mu$,
  $\xi$ may be thought of as $\xi = \frac{1}{2}(\rho-\sigma)$ for
  orthogonal states $\rho$, $\sigma$.
\end{lemma}

\begin{proof}
The optimisation problems are an immediate consequence of the definitions, and 
we already argued that $\mu_1(\mathbb{M}) = 1$. To lower bound $\lambda_1(\mathbb{M})$
we proceed as follows:
Given any $\xi$ of trace norm $1$, we can write it as 
\[
  \xi = (1-p)\rho - p\sigma
      = (1-p)(\rho-\sigma) + (1-2p)\sigma
      = 2(1-p)\xi_0 + (1-2p)\sigma,
\]
with orthogonal states $\rho$ and $\sigma$, and $\xi_0 = \frac{1}{2}(\rho-\sigma)$.
W.l.o.g.~$0 \leq p \leq 1/2$, otherwise use $-\xi$.
Now let $X_0 \in \mathbb{M}$ be optimal for $\xi_0$,
i.e.~$\| \xi_0 \|_{\mathbb{M}} = \tr (\xi_0 X_0)$, and test $\xi$
with $X=(\1+X_0)/2 \in \mathbb{M}$. Note $X \geq 0$, so
\[\begin{split}
  \| \xi \|_{\mathbb{M}}  =    \tr (\xi X)
                         &=    (1-p)\tr (\xi_0 X) + \frac{1-2p}{2}\tr (\sigma X)    \\
                         &=    (1-p)\tr(\xi_0 X_0) + \frac{1-2p}{2}\tr (\sigma X)  \\
                         &\geq \frac{1}{2}\tr (\xi X_0)                              
                          =    \frac{1}{2}\| \xi_0 \|_{\mathbb{M}}
                          \geq \frac{1}{2}\lambda(\mathbb{M}),
\end{split}\]
concluding the proof.
\end{proof}

What is the relation of the constants of domination for different sets $\mathbb{M}$ and 
$\mathbb{M}'$? Clearly, if $\mathbb{M} \subseteq \mathbb{M}'$, then
$\lambda(\mathbb{M}) \leq \lambda(\mathbb{M}')$ and
$\mu(\mathbb{M}) \leq \mu(\mathbb{M}')$. More interesting relations
are obtained by using the convex structure.
For this purpose we look at convex combinations of POVMs in the sense
of direct sums as follows. For POVMs $(M_k)$ and $(N_\ell)$,
and a real $0\leq p \leq 1$, we denote by
$p (M_k)_k \oplus (1-p) (N_\ell)_\ell$
the POVM consisting of the $m+n$ elements
\[
  p M_1, \ldots, p M_m, (1-p) N_1, \ldots (1-p) N_n.
\]
If the associated CPTP maps of the two original POVMs are ${\cal M}$
and ${\cal N}$, then the direct sum convex combination has associated CPTP map
\[
  p{\cal M} \oplus (1-p){\cal N}: \xi \longmapsto \sum_{k=1}^m \proj{k} p\tr(\xi M_k)
                                      + \sum_{\ell=1}^n \proj{m\!+\!\ell} (1-p)\tr(\xi N_\ell).
\]
If we have two sets of POVMs, $\mathbf{M}$ and $\mathbf{N}$, then their
direct sum convex combination is defined naturally as
\[
  p\mathbf{M} \oplus (1-p)\mathbf{N} 
      = \bigl\{ p (M_k)_k \oplus (1-p) (N_\ell)_\ell : (M_k) \in \mathbf{M},\ 
                                                       (N_\ell) \in \mathbf{N} \bigr\}.
\]

More generally, we can look at convex combinations of any finite or
even countable number of POVMs and sets of POVMs. These constructions
have a straightforward operational interpretation: implementing
$p (M_k)_k \oplus (1-p) (N_\ell)_\ell$ means tossing a biased
coin, with $p$ being the probability of heads, then measuring $(M_k)$
if heads showed, and $(N_\ell)$ for tails. The coin toss is part of 
the measurement result.

\begin{lemma}
  \label{lemma:convex-comb}
  Let $\mathbf{M}_i$ be sets of POVMs and $p_i\geq 0$ probabilities,
  and $\mathbf{N} = \bigoplus_i p_i\mathbf{M}_i$. Denote the corresponding
  sets of operators $\mathbb{M}_i$ and $\mathbb{N}$.
  Then,
  \[
    \mathbb{N} = \sum_i p_i \mathbb{M}_i,
  \]
  and consequently
  \[
    \lambda(\mathbb{N}) \geq \sum_i p_i \lambda(\mathbb{M}_i),\quad
    \mu(\mathbb{N}) \leq \sum_i p_i \mu(\mathbb{M}_i).
  \]
\end{lemma}
\begin{proof}
  The first relation is by inspection. For the inequalities, note that
  since we have
  \[
    \lambda(\mathbb{M}_i) [-\1;\1]/_{\RR\scriptidop} \subseteq \mathbb{M}_i/_{\RR\scriptidop} 
                                                \subseteq \mu(\mathbb{M}_i) [-\1;\1]/_{\RR\scriptidop},
  \]
  we clearly get
  \[
    \sum_i p_i \lambda(\mathbb{M}_i) [-\1;\1]/_{\RR\scriptidop} \subseteq \sum_i p_i \mathbb{M}_i/_{\RR\scriptidop}
                                                 \subseteq \sum_i p_i \mu(\mathbb{M}_i) [-\1;\1]/_{\RR\scriptidop}.
  \]
\end{proof}

In particular, since $[-\1;\1]$ is invariant under unitary conjugation,
i.e.~$U [-\1;\1] U^\dagger = [-\1;\1]$, the constants of domination also have
this invariance and so we obtain immediately
\begin{proposition}
  \label{prop:symmetrise}
  For a probability measure ${\rm d}p(U)$ on the unitary group on ${\cal H}$,
  and any symmetric, full dimensional convex body $\pm\1 \in \mathbb{M} \subseteq [-\1;\1]$,
  \begin{align*}
    \lambda\left( \int {\rm d}p(U) U\mathbb{M}U^\dagger \right) &\geq \lambda(\mathbb{M}), \\
    \mu\left( \int {\rm d}p(U) U\mathbb{M}U^\dagger \right)     &\leq \mu(\mathbb{M}).
  \end{align*}
  In other words: symmetrisation makes $\mathbb{M}$ ``look more like $[-\1;\1]$''.
  \qed
\end{proposition}

\section{Single POVMs}
\label{sec:single}
Let us look now at the constants of domination $\lambda$ and $\mu$
in the case that $\mathbf{M}$ consists of a single, informationally-complete, POVM.
Let ${\cal M}$  be the CPTP map associated with the POVM.
With slight abuse of notation we denote the constants of
domination $\lambda({\cal M})$ and $\mu({\cal M})$, and the associated
norm $\| \cdot \|_{{\cal M}}$.

\subsection{Uniform POVM}
As a consequence of Proposition~\ref{prop:symmetrise}, we arrive at the following theorem.
\begin{theorem}
  \label{thm:uniform}
  The supremum of $\lambda({\cal M})$, as well as the infimum of $\mu({\cal M})$,
  over all POVMs in dimension $d$ is attained by the uniform (unitary
  invariant POVM),
  \[
    \bigl( d\proj{\psi} {\rm d}\psi \bigr), \text{ with the normalised uniform distribution }
                                            {\rm d}\psi \text{ on unit vectors.}
  \]
  Furthermore, denoting the CPTP map associated with it by ${\cal U}$,
  \begin{align}
    \label{eq:lambda-uniform}
    \lambda({\cal U}) &= \min_{1\leq a\leq d/2 \atop b=d-a}
                            1 - \frac{1}{d}\sum_{k=0,\ldots a-1 \atop \ell=0,\ldots b-1}
                                 \Bigl(\frac{a}{d}\Bigr)^k \Bigl(\frac{b}{d}\Bigr)^\ell {k+\ell \choose k}
                       = \sqrt{\frac{1}{d}} \left(\sqrt{\frac{2}{\pi}} \pm o(1) \right) \\
    \label{eq:mu-uniform}
    \mu({\cal U})     &= \frac{1}{2}.
  \end{align}
\end{theorem}
\begin{proof}
We have already argued the supremum and infimum, so we are left to prove
eqs.~(\ref{eq:lambda-uniform}) and (\ref{eq:mu-uniform}). Note that for
any operator $\xi$,
\begin{equation}
  \label{eq:U-bias}
  \| {\cal U}(\xi) \|_1 = d \int {\rm d}\psi |\tr (\psi\xi) |.
\end{equation}

Note that since $\xi$ is Hermitian, we may again take $\xi = (1-p) \rho - p \sigma$ for orthogonal operators 
$\rho$ and $\sigma$.
For eq.~(\ref{eq:lambda-uniform}), we then have by the unitary invariance of the
uniform POVM and the triangle inequality, $\lambda({\cal U})$ is
attained as $\| {\cal U}(\xi) \|_1$ for an operator of the form
\[
  \xi = \frac{1}{2a}P - \frac{1}{2b}Q,\quad\text{with a projector }P\text{ of rank }a,
                                          \text{ and }Q=\1-P,\ b=d-a,
\]
where we may even take $P$ to be the projector onto the subspace spanned
by the first $a$ computational basis vectors (again invoking unitary
invariance).
For this choice of operator, according to eq.~(\ref{eq:U-bias}), and letting $p:=a/d$,
\[
  \| {\cal U}(\xi) \|_1
         = d \int {\rm d}\psi \left| \frac{1}{2a}\sum_{j=1}^a |\psi_j|^2 
                                     - \frac{1}{2b}\sum_{j=a+1}^d |\psi_j|^2 \right|
         = 1 - \frac{1}{d}\sum_{k=0,\ldots a-1 \atop \ell=0,\ldots b-1}
                                         p^k (1-p)^\ell {k+\ell \choose k},
\]
by Lemma~\ref{lemma:integral} in Appendix~\ref{sec:integral}.
It is quite natural to conjecture that the minimal choice of ranks is
$a=\lfloor d/2 \rfloor$ and $b=\lceil d/2 \rceil$. In this case we have 
\begin{equation}
  \label{eq:conjecture}
  \| {\cal U}(\xi) \|_1 = 1 - \frac{1}{d}\sum_{k=0,\ldots \lfloor d/2 \rfloor - 1 
                                                \atop \ell=0,\ldots \lceil d/2 \rceil - 1}
                                           \left(\frac{\lfloor d/2 \rfloor}{d}\right)^k
                                           \left(\frac{\lceil d/2 \rceil}{d}\right)^\ell 
                                           {k+\ell \choose k}         
                        = \sqrt{\frac{2}{\pi d}} \pm O\left(\frac{1}{d}\right),
\end{equation}
for large $d$. 
The analysis of the asymptotics is elementary but lengthy, and is here restricted
to a few hints: We lose only terms of order $O(1/d)$ by focusing on
even $d$, for which the formula evaluates to
\[
  \lambda({\cal U}) = 1 - \frac{1}{d}\sum_{k,\ell=0}^{d/2-1} 2^{-k-\ell} {k+\ell \choose k} 
                    = \frac{1}{d} \sum_{k=0}^{d/2-1} 2^{-2k} {2k \choose k},
\]
where we have used the following identity from Lemma~\ref{lemma:binomial},
proved by induction on $k$:
\begin{equation}
  \label{eq:neat-identity}
  \sum_{\ell=0}^k 2^{-k-\ell} {k+\ell \choose \ell} = 1.
\end{equation}
Then a simple application of Stirling's formula (with explicit error bounds)
yields eq.~(\ref{eq:conjecture}).

However, we have not been able to prove that this is indeed the minimum. Instead,
we follow a different route: From the proof of Lemma~\ref{lemma:integral} in Appendix~\ref{sec:integral},
we observe that for general $a$ and $b$,
\[
  \| {\cal U}(\xi) \|_1
         = \EE \left| \frac{1}{2a}\sum_{j=1}^a X_j - \frac{1}{2b}\sum_{j=a+1}^d X_j \right|,
\]
with independent $X_j \geq 0$, each distributed according to a rescaled $\chi^2_2$ law.
By definition, their expectation and variance are $\EE X_j = 1$ and 
$\operatorname{Var} X_j = 1$, respectively (also, all higher moments are finite).
Thus, by the central limit theorem
\[
  \frac{1}{2a}\sum_{j=1}^a X_j \approx Y_0 \sim {\cal N}\left( 1, \frac{1}{4a} \right),\quad
  \frac{1}{2b}\sum_{j=a+1}^d X_j \approx Y_1 \sim {\cal N}\left( 1, \frac{1}{4b} \right),
\]
where $Y_0$ and $Y_1$ are normal distributed with means $\mu$ and variance $\nu$ as indicated by ${\cal N}(\mu, \nu)$,
and the approximation signs indicate convergence in probability as 
$a,b \rightarrow \infty$. (Note that since the third moment of the $X_j$
is finite, this convergence is uniform in $a$ and $b$, thanks to the
Berry-Ess\'{e}en theorem which bounds the rate of convergence in the
central limit theorem -- see e.g.~\cite{Bolthausen}.)

Since $Y_0 - Y_1 =: Z \sim {\cal N}\left( 0, \frac{1}{4a}+\frac{1}{4b} \right)$,
we obtain asymptotically
\[
  \| {\cal U}(\xi) \|_1 \sim \EE |Z| = \sqrt{\frac{1}{4a}+\frac{1}{4b}}
                                            \frac{1}{\sqrt{2\pi}} \int_{-\infty}^\infty {\rm d}x |x| e^{-x^2/2}
                                     = \sqrt{\frac{2}{\pi}} \sqrt{\frac{1}{4a}+\frac{1}{4b}},
\]
which is minimized for $a = b = d/2$, yielding
$\lambda({\cal U}) \sim \sqrt{\frac{2}{\pi d}}$, as advertised.

For eq.~(\ref{eq:mu-uniform}), note that by the triangle inequality,
$\mu({\cal M})$ of any POVM ${\cal M}$ is attained as $\| {\cal M}(\xi) \|_1$ for
an extremal traceless $\xi$ such that $\| \xi \|_1 = 1$. These are
easily seen to be of the form $\xi = \frac{1}{2}\proj{\phi_1} - \frac{1}{2}\proj{\phi_2}$
for orthogonal pure state vectors $\ket{\phi_1}$, $\ket{\phi_2}$. By
unitary invariance of the uniform POVM, any such $\xi$ will in fact
yield the same value, so we may take $\xi = \frac{1}{2}\proj{1}-\frac{1}{2}\proj{2}$,
so that by eq.~(\ref{eq:U-bias}),
\[
  \mu({\cal U}) = \| {\cal U}(\xi) \|_1
                = \frac{d}{2} \int {\rm d}\psi \bigl| |\psi_1|^2 - |\psi_2|^2 \bigr|
                = \frac{1}{2},
\]
once more by Lemma~\ref{lemma:integral} in Appendix~\ref{sec:integral},
applied with $a=b=1$.
\end{proof}

\noindent
Note that in terms of the bias the above translates to
\[
  \frac{1}{2}\| \rho - \sigma \|_1 
      \geq \| \rho - \sigma \|_{\cal M} 
      \geq \sqrt{\frac{1}{d}} \left(\sqrt{\frac{2}{\pi}} - o(1) \right) \| \rho - \sigma \|_1.
\]

\subsection{Almost optimal performance of $4$-designs}
\label{subsec:t-design}
The results of the previous section provide the motivation to look at
POVMs made from $t$-designs, as these are structures approximating
the full random POVM better and better as $t \rightarrow \infty$.
We thus intuitively expect to obtain a similar value for $\lambda$
as we obtained for the random POVM for larger $t$.
Recall the following

\begin{definition}
A (weighted) spherical $t$-design is an ensemble
$( p_k, P_k )_{k=1}^n$ of $1$-dimensional projectors $P_k$ and
probabilities $p_k$ such that
\[
  \sum_k p_k P_k^{\ox t} = \frac{1}{{d+t-1 \choose t}} P^{(t)}_{\rm sym}
                           = \frac{1}{N(d,t)}\sum_{\pi \in S_t} U_\pi,
\]
with the projector $P^{(t)}_{\rm sym}$ onto the completely symmetric
subspace of $(\CC^d)^{\ox t}$, which has dimension
${d+t-1 \choose t}$. It can be expressed, by Schur duality,
via the representation of the symmetric group (as this subspace is
an irrep of multiplicity $1$), as the sum of the permutation representations
$U_\pi$, which is the permutation of the $t$ tensor
factors by the permutation $\pi\in S_t$. Since $\tr (U_\pi) = d^{c(\pi)}$
with the number $c(\pi)$ of cycles of the permutation $\pi$,
we get the normalisation factor
$N(d,t) = \sum_{\pi\in S_t} d^{c(\pi)} = (d+t-1)\cdots(d+1)d$.
\end{definition}

Note that the random POVM $\bigl( d\proj{\psi} {\rm d}\psi \bigr)$
is an $\infty$-design.
We call a $t$-design \emph{proper} if all the probabilities
are equal, $p_k = 1/n$. Note that any $t$-design is automatically
also a $t'$-design for all $t' < t$. In particular,
$\sum_k p_k P_k = \frac{1}{d}\1$, so it makes sense to associate a
POVM with every $t$-design of the form
\[
  (M_k)_{k=1}^n,\quad \text{with }M_k = d p_k P_k,
\]
which, as before, we also call a (weighted or proper) $t$-design.
We shall consider the CPTP map ${\cal M}$ associated with such a $t$-design POVM.

It turns out that $4$-designs already achieve essentially the same
bias as the isotropic POVM. This was
discovered by Ambainis and Emerson~\cite{Ambainis:Emerson}, who showed, invoking
a beautiful moment inequality by Berger, that
\begin{equation}
  \label{eq:AE}
  \| \rho-\sigma \|_{{\cal M}} \geq \frac{1}{3}\| \rho-\sigma \|_2
                               \geq \frac{1}{3\sqrt{d}} \| \rho-\sigma \|_1.
\end{equation}
We briefly review their argument, including the Berger inequality, as
we need to return to this later on in Section~\ref{sec:local-POVM}.

\begin{lemma}[Berger~\cite{Berger}]
  \label{lemma:berger}
  For a real random variable $S$,
  \[
    \EE |S| \geq \frac{(\EE S^2)^{3/2}}{(\EE S^4)^{1/2}}.
  \]
\end{lemma}
\begin{proof}
  That is just H\"older's inequality, which states that for
  real random variables $f$ and $g$, and $\frac{1}{p}+\frac{1}{q} = 1$,
  \[
    \EE( fg ) \leq \left(\EE|f|^p\right)^{1/p} \left(\EE|g|^q\right)^{1/q}.
  \]
  Here it is applied with $f = |S|^{2/3}$, $g = |S|^{4/3}$
  and $p=3/2$, $q=3$.
\end{proof}

\begin{proof}[of eq.~(\ref{eq:AE}) -- see~\cite{Ambainis:Emerson}]
For traceless $\xi$, consider the random variable $S$ which takes
value $d \tr (\xi P_k)$ with probability $p_k$. Then clearly
$\EE |S| = \| {\cal M}(\xi) \|_1$, and Berger's inequality can be used.
The moments are easy calculations, using the fact that the POVM is
a $4$-design. First, the second moment,
\[\begin{split}
  \EE S^2 &= \sum_k p_k d^2 \bigl( \tr (\xi P_k) \bigr)^2 \\
          &= \sum_k p_k d^2 \tr\bigl( (\xi\ox\xi)(P_k \ox P_k) \bigr) \\
          &= d^2 \tr\left( (\xi\ox\xi)\frac{2}{d(d+1)}P^{(2)}_{\rm sym} \right) \\
          &= \frac{d^2}{d(d+1)} \tr\bigl( (\xi\ox\xi)(\1+F) \bigr)
           = \frac{d}{d+1}\tr (\xi^2),
\end{split}\]
where $F$ is the swap operator, corresponding to the transposition $(12)$,
and we have made use of $\tr(\xi)=0$.
Similarly,
\[\begin{split}
  \EE S^4 &= \sum_k p_k d^4 \bigl( \tr (\xi P_k) \bigr)^4 \\
          &= \sum_k p_k d^4 \tr\bigl( (\xi\ox\xi\ox\xi\ox\xi)(P_k \ox P_k \ox P_k \ox P_k) \bigr) \\
          &= d^4 \tr\left( \xi^{\ox 4}\frac{24}{d(d+1)(d+2)(d+3)}P^{(4)}_{\rm sym} \right) \\
          &= \frac{d^4}{d(d+1)(d+2)(d+3)} \sum_{\pi\in S_4 } \tr \bigl(\xi^{\ox 4} U_\pi\bigr) \\
          &= \frac{d^3}{(d+1)(d+2)(d+3)} \bigl( 3(\tr(\xi^2))^2 + 6 \tr(\xi^4) \bigr) 
           \leq \left(\frac{d}{d+1}\right)^3 \, 9(\tr(\xi^2))^2,
\end{split}\]
where in the last line we have made use of $\tr(\xi)=0$ to take care of all
permutations with a fixed point.
Thus,
\[
  \| {\cal M}(\xi) \|_1 = \EE |S| \geq \frac{1}{3}\sqrt{\tr(\xi^2)}
                                  =    \frac{1}{3}\| \xi \|_2 \geq \frac{1}{3\sqrt{d}}\| \xi \|_1.
\]
In other words: $\lambda({\cal M}) \geq 1/(3\sqrt{d})$.
\end{proof}

It is not known how to construct spherical $4$-designs efficiently in general
though Caratheodory's Theorem tells us that there must exist a weighted
$4$-design of cardinality at most $1+{d+3 \choose 4}^2$. 
Constructions are known for a real vector space of small dimensions~\cite{hs:designs}.
Ambainis
and Emerson~\cite{Ambainis:Emerson} construct approximate $4$-designs
which perform almost as good as eq.~(\ref{eq:AE}).

\subsection{Performance of $2$-designs}
\label{subsec:2-design}
Unfortunately, we cannot give any bounds for the bias for
$3$-design POVMs, but here we show how to bound
it for $2$-designs. Consider first a proper $2$-design with associated
POVM $(M_k = \frac{d}{n} P_k)_{k=1}^n$. I.e.,
\[
  \frac{1}{n}\sum_k P_k \ox P_k = \frac{1}{d(d+1)}(\1+F) = \frac{2}{d(d+1)} P^{(2)}_{\rm sym},
\]
with the projector $P^{(2)}_{\rm sym}$ onto the symmetric subspace of $\CC^d \ox \CC^d$
and the swap operator $F$. Such POVMs are always tomographically complete -- this
will also follow from the theorem below.

An example of a 2-design is a complete set of $d+1$ mutually unbiased bases,
which are known to exist if the dimension $d$ is a prime power~\cite{WoottersFields,BBRV}).
Let
\[
  \bigl\{ (\ket{\psi^b_s})_{s=1\ldots d} : b = 0,\ldots,d \bigr\},
\]
be the basis vectors of the $d+1$ mutually unbiased bases, where $\ket{\psi^b_s}$ is the $s$-th basis
vector of the $b$-th basis. Then the set of basis state projectors $P^b_s = \proj{\psi^b_s}$ 
forms a proper spherical 2-design~\cite{kr:mubDesign}.
It is conjectured that in all dimensions there exist spherical $2$-designs
with the minimum number $n=d^2$ of elements~\cite{sic-povm}, giving rise to so-called
\emph{symmetric informationally complete} (SIC) POVMs. These are only
known to exist up to dimension $d=45$~\cite{sic-povm} by numerical results, and
for even fewer dimensions up to $d=19$ by mathematical construction.
Zauner's conjecture states that
in every dimension there exists a SIC-POVM of a particularly beautiful group
symmetric form~\cite{zauner}. We refer to~\cite{appleby:sic-povm,flammia} for more information.

As before, we look at the associated CPTP map,
\[
  {\cal M}: \xi \longmapsto \sum_{k=1}^n \proj{k} \tr(\xi M_k)
                            = \frac{d}{n} \sum_k \proj{k} \tr(\xi P_k).
\]
Our objective is to prove the relation.
\begin{theorem}
  \label{thm:2-design}
  For any traceless Hermitian operator $\xi$,
  \begin{equation}
    \label{eq:certainty}
    \| {\cal M}(\xi) \|_1 \geq \frac{1}{2}\frac{1}{d+1} \| \xi \|_1.
  \end{equation}
  In other words, for any proper 2-design POVM as above,
  $\lambda({\cal M}) \geq \frac{1}{2}\frac{1}{d+1}$.
\end{theorem}
\begin{proof}
Since this is a homogeneous relation, we may w.l.o.g.~assume that $\| \xi \|_1 = 2$,
meaning that we can write $\xi = \rho-\sigma$ with two orthogonal
density operators $\rho$ and $\sigma$. Thus, what we need to
show is $\| {\cal M}(\rho) - {\cal M}(\sigma) \|_1 \geq \frac{1}{d+1}$.

For this, we use Proposition~\ref{prop:l1-ineq} in
Appendix A, ineq.~(\ref{eq:conj}),
for the vectors $\vec{p}$ and $\vec{q}$ defined as
\[
  p_k = \tr (\rho M_k) = \frac{d}{n} \tr (\rho P_k),\quad
  q_k = \tr (\sigma M_k) = \frac{d}{n} \tr (\sigma P_k).
\]
Namely,
\[\begin{split}
  \| {\cal M}(\rho) - {\cal M}(\sigma) \|_1
       &=    \| \vec{p} - \vec{q} \|_1                                  \\
       &\geq 1 - n \sum_k \frac{d^2}{n^2}(\tr (\rho P_k))(\tr (\sigma P_k)) \\
       &=    1 - d^2 \frac{1}{n}\sum_k \tr\bigl( (\rho\ox\sigma)(P_k \ox P_k) \bigr).
\end{split}\]
Now, the last sum can be evaluated as follows, using the property of
spherical $2$-design:
\[\begin{split}
  \frac{1}{n}\sum_k \tr(\rho P_k \sigma P_k)
       &= \frac{1}{n} \sum_k \tr \L( (P_k \ox P_k) (\rho \ox \sigma) \R) \\
       &= \frac{1}{d(d+1)}\tr \L( (\1+F)(\rho \ox \sigma) \R) \\
       &= \frac{1}{d(d+1)}(\tr(\rho)\tr(\sigma) + \tr(\rho\sigma))
        = \frac{1}{d(d+1)},
\end{split}\]
Inserting this above, we conclude
\[
  \| {\cal M}(\rho) - {\cal M}(\sigma) \|_1
        \geq 1 - d^2 \frac{1}{d(d+1)} = \frac{1}{d+1},
\]
as advertised.
\end{proof}

\begin{corollary}
  \label{cor:2-design}
  For a POVM which is a weighted $2$-design, and associated map
  ${\cal M}$, the conclusion of Theorem~\ref{thm:2-design} still
  holds: $\lambda({\cal M}) \geq \frac{1}{2}\frac{1}{d+1}$.
\end{corollary}
\begin{proof}
  The idea is to break down the probabilities $p_k$ into smaller
  but approximately equal values. This increases the number of
  outcomes of the POVM, but makes it be approximated better and better
  by a proper $2$-design, to which we can apply Theorem~\ref{thm:2-design}.
  
  In detail, assume that our weighted $2$-design is discrete, with $n$
  elements; choose an integer $N \gg 1$, and for each $k$
  let $N_k = \lfloor N p_k \rfloor$ and $\epsilon_k = N p_k - N_k$.
  Define a new weighted $2$-design with the same projectors $P_{k\ell} = P_k$
  and ``uniformised'' weights
  \[
    \beta_{k\ell} = \begin{cases} \epsilon_k/N & \text{ for }\ell=0, \\
                                  1/N          & \text{ for }\ell=1,\ldots,N_k.
                    \end{cases}
  \]

  Then, applying the same proof as in Theorem~\ref{thm:2-design} to this
  refined $2$-design (which has $N+n$ outcomes), we get
  \[\begin{split}
    \| {\cal M}(\rho) - {\cal M}(\sigma) \|_1
       &=    \| \vec{p} - \vec{q} \|_1                                                   \\
       &\geq 1 - (N+n) \sum_{k\ell} \beta_{k\ell}^2 d^2 (\tr (\rho P_k))(\tr (\sigma P_k))     \\
       &\geq 1 - d^2 \frac{N+n}{N} \sum_{k\ell} \beta_{k\ell} \tr\bigl( (\rho\ox\sigma)(P_k \ox P_k) \bigr) \\
       &=    1 - \frac{d^2}{d(d+1)} \frac{N+n}{N} \tr \bigl[ (\1+F)(\rho \ox \sigma) \bigr]  \\
       &=    1 - \frac{d}{d+1}\left( 1+\frac{n}{N} \right)
        \rightarrow \frac{1}{d+1},
  \end{split}\]
  where we have used $\beta_{k\ell} \leq 1/N$ in the third line.
\end{proof}

\bigskip

Note that the factor of $1/(d+1)$ in the bound~(\ref{eq:certainty})
is essentially best possible (up to a constant independent of $d$),
as the example of $d+1$ mutually unbiased bases shows. Indeed, if
the two states $\rho$ and $\sigma$ are distinct elements of one
of the bases, then the measured output distributions for all
the $d$ other bases are the same, namely uniform, while in their
proper basis the trace distance remains $2$, so
\(
   \| {\cal M}(\rho) - {\cal M}(\sigma) \|_1 = \frac{2}{d+1},
\)
and hence $\lambda({\cal M}) \leq \frac{1}{d+1}$.

Similarly, for a SIC-POVM with $d^2$ operators $\left( \frac{1}{d}P_k \right)$
it is easily verified that two states from the POVM, i.e.~for
instance $\rho = P_1$ and $\sigma = P_2$, have trace norm difference
$\| \rho-\sigma \|_1 = \frac{2d}{d+1}$, while
\(
  \| {\cal M}(\rho) - {\cal M}(\sigma) \|_1 = \frac{2}{d+1},
\)
so $\lambda({\cal M}) \leq \frac{1}{d}$.

\section{Local POVMs}
\label{sec:local-POVM}
Consider now a multipartite system
${\cal H} = {\cal H}_1 \ox {\cal H}_2 \ox \cdots \ox {\cal H}_n$,
of local Hilbert spaces ${\cal H}_j$ of dimension $d_j$. (The total
space's dimension is denoted $D = d_1 d_2 \cdots d_n$ in this section.)
This partition suggests various classes of POVMs due to restrictions
of locality. For instance, let $\textbf{LO}$ be the class of
all \emph{local operations}, i.e.~tensor product measurements:
\[
  \textbf{LO} = \left\{ \bigl( M^{(1)}_{k_1} \ox \cdots \ox M^{(n)}_{k_n} \bigr) : 
                            (M^{(j)}_{k_j}) \text{ POVM on } {\cal H}_j \right\}.
\]
More generally, $\textbf{LOCC}$ is the class of measurements that can be
implemented by local operations and classical communication between the
parties. $\textbf{SEP}$ are the separable POVMs, i.e.
\[
  \textbf{SEP} = \left\{ \bigl( M^{(1)}_{k} \ox \cdots \ox M^{(n)}_{k} \bigr) : 
                            M^{(j)}_k \geq 0,\ \sum_k M^{(1)}_{k} \ox \cdots \ox M^{(n)}_{k} = \1 \right\}.
\]
Finally, there is the class of all measurements positive partial transpose
(PPT) operators: denoting the transpose operation (with respect to any
basis) by $T$, it is
\[
  \textbf{PPT} = \left\{ \bigl( M_k \bigr) \text{ POVM}:\  
                         \forall k \forall I\subset[n]\ 
         \left( \bigotimes_{i\in I} T \ox \bigotimes_{i\not\in I} \id \right) M_k \geq 0 \right\},
\]
i.e.~all POVM elements have to be PPT with respect to every bipartition
of the $n$-party system.

Quite evidently,
\[
  \textbf{LO} \subset \textbf{LOCC} \subset \textbf{SEP} \subset \textbf{PPT},
\]
and all inclusions are well-known to be strict, at least if the
dimension is large enough. The corresponding symmetric convex bodies of
operators are denoted
\[
  \mathbb{LO} \subset \mathbb{LOCC} \subset \mathbb{SEP} \subset \mathbb{PPT}.
\]

These are interesting examples of POVM classes since we know due to
so-called quantum data hiding~\cite{data-hiding1,data-hiding2,Matthews:Winter}
that $\| \xi \|_{\mathbb{M}}$ for them can be much smaller than $\| \cdot \|_1$.
Indeed, it was shown in these references that in a bipartite system
$\CC^d \ox \CC^d$, the states $\sigma = \frac{\scriptidop+F}{d(d+1)}$ and 
$\alpha = \frac{\scriptidop-F}{d(d-1)}$, i.e.~the (normalised) projectors onto
the symmetric and antisymmetric subspace, respectively, obey
\[
  \left\| \frac{1}{2}\rho - \frac{1}{2}\sigma \right\|_{\mathbb{PPT}} = \frac{2}{d+1}.
\]
(In~\cite{data-hiding2} more general statements of this type for $n$-partite
systems can be found.) Consequently, $\lambda(\mathbb{PPT}) \leq \frac{2}{d+1}$.
The next result shows that this bound is not very far from the truth:

\begin{lemma}
  \label{lemma:sep-bound}
  For any operator $\xi$ on an $n$-partite system,
  \[
    \| \xi \|_{\mathbf{SEP}} \geq \frac{2}{2^{n/2}} \| \xi \|_2
                             \geq \frac{2}{2^{n/2}} \frac{1}{\sqrt{D}}\| \xi \|_1.
  \]
  In particular, $\lambda(\mathbb{SEP}) \geq \frac{2}{2^{n/2}} \frac{1}{\sqrt{D}}$;
  for a bipartite system, we find $\lambda(\mathbb{SEP}) \geq \frac{1}{\sqrt{D}}$.
\end{lemma}
\begin{proof}
	Gurvits and Barnum~\cite{BarnumGurvits} 
	have shown that for a bipartite system,
	within the set of Hermitian operators, the unit
	ball of the Hilbert-Schmidt norm centred on the identity operator
	contains only separable operators. More generally
	they proved in an $n$-partite system, that the ball of radius
	$2^{1-n/2}$ around the identity is fully separable~\cite{BarnumGurvits}.
	
	It follows immediately that all the POVMs in the set 
	$\left\{ \L( M, \1 - M \R) : \| 2M - \1 \|_2 \leq 2^{1-n/2} \right\}$
	are separable. It is easy to see that the corresponding symmetric convex
	body (see Lemma \ref{lemma:2-outcome}) is the ball of radius $2^{1-n/2}$ in the Hilbert-Schmidt norm
	around the origin and so this is a subset of $\mathbb{SEP}$.
	
	From this inclusion, and the fact that the Hilbert-Schmidt norm is self-dual,
	\[
		\|\xi\|_{\mathbf{SEP}} = \max_{M \in \mathbb{SEP}}
		                             \tr\L( M \xi \R) \geq \max_{\|M\|_2 \leq 2^{1-n/2}} \tr\L( M \xi \R) 
		                       = \frac{2}{2^{n/2}} \|\xi\|_2,
	\]
	concluding the proof, if we recall $\|\xi\|_1 \leq \sqrt{D} \|\xi\|_2$.
\end{proof}

We now come to the main technical result of the present section, showing that
this order of magnitude goes through all the way to $\mathbb{LO}$,
indeed, a particular tensor product POVM on a bipartite system is already
almost as good as the class of all separable POVMs, in terms
of the constant of domination. Note that Proposition~\ref{prop:symmetrise}
gives us the local POVM with the largest $\lambda$: namely, by
symmetrising over all unitaries $U = U_A \ox U_B$, drawn from the
product of the local Haar measures, we find that for any tensor product
POVM ${\cal M}_A \ox {\cal N}_B$, we have 
$\lambda\bigl( {\cal U}_A\ox{\cal U}_B \bigr) \geq \lambda\bigl( {\cal M}_A\ox{\cal N}_B \bigr)$.

\begin{theorem}
  \label{thm:per-aspera-ad-astram}
  For any two states $\rho$ and $\sigma$ on a bipartite Hilbert space
  ${\cal H}_A \ox {\cal H}_B$, let $\xi = \rho-\sigma$. Then,
  \[
    \| \xi \|_{{\cal U}_A\ox{\cal U}_B} 
                            \geq \frac{1}{\sqrt{153}} \| \xi \|_2
                            \geq \frac{1}{\sqrt{153 D}} \| \xi \|_1,
  \]
  where $D= d_A d_B$ is the Hilbert space dimension, and ${\cal U}_A$
  and ${\cal U}_B$ are the CPTP maps of the isotropic POVMs on ${\cal H}_A$
  and ${\cal H}_B$, respectively.
  Consequently, $\lambda\bigl( {\cal U}_A\ox{\cal U}_B \bigr) \geq 1/\sqrt{153 D}$.
\end{theorem}
\begin{proof}
  We do exactly the same as in Subsection~\ref{subsec:t-design},
  only that we have now a POVM on ${\cal H}_A \ox {\cal H}_B$ of the form
  \[
    \bigl(  D {\rm d}\varphi{\rm d\psi}  \ketbra{\varphi}{\varphi}\ox\ketbra{\psi}{\psi} \bigr),
  \]
  so $S$ is the variable
  \[
    S = D \tr ((\Pvphi \ox \Ppsi)\xi),
  \]
  and the bias of the estimation based on the outcome is $\EE |S|$, as before in
  Subsection~\ref{subsec:t-design}.

  We use Berger's inequality, Lemma~\ref{lemma:berger} again, for which
  we need the second and fourth moment. Because now we randomise
  independently over ${\cal H}_A$ and ${\cal H}_B$, we get 
  \begin{align*}
    \EE S^2 &= \frac{2^2 d_A^2 d_B^2}{d_A(d_B+1)d_B(d_B+1)} 
                \tr\bigl( (\Pi^{AA}_{\rm sym}\ox\Pi^{BB}_{\rm sym})(\xi^{AB} \ox \xi^{AB}) \bigr), \\
    \EE S^4 &= \frac{24^2 d_A^4 d_B^4}{d_A(d_A+1)(d_A+2)(d_A+3)d_B(d_B+1)(d_B+2)(d_B+3)}       \\
            &\phantom{====}
                 \times\tr\bigl( (\Pi^{AAAA}_{\rm sym}\ox\Pi^{BBBB}_{\rm sym})
                           (\xi^{AB} \ox \xi^{AB} \ox \xi^{AB} \ox \xi^{AB}) \bigr),
  \end{align*}
  where the superscripts remind of the systems these operators act on.

  Expanding the projectors into the permutations of two, respectively four, elements, we get
  \begin{equation}
    \label{eq:UU-2nd-moment}
    \EE S^2 =  \frac{d_A d_B}{(d_A+1)(d_B+1)} \left( \tr (\xi_A^2) + \tr (\xi_B^2) + \tr (\xi^2) \right),
  \end{equation}
  where $\xi_A = \tr_B (\xi)$ and $\xi_B = \tr_A (\xi)$, because we get terms with $\1^{AA} \ox \1^{BB}$,
  $\1^{AA} \ox F^{BB}$, $F^{AA} \ox \1^{BB}$ and $F^{AA} \ox F^{BB}$.

  The fourth moment is considerably more complex: looking at
  \begin{equation}
    \label{eq:UU-4th-moment}
    \EE S^4 = \frac{d_A^3 d_B^3}{(d_A+1)(d_A+2)(d_A+3)(d_B+1)(d_B+2)(d_B+3)} 
                \sum_{\pi,\sigma \in S_4} \tr\bigl( (U_\pi^{AAAA} \ox U_\sigma^{BBBB}) \xi^{\ox 4} \bigr),
  \end{equation}
  we see that we need to calculate -- or at least reasonably upper bound --
  the trace terms $\tr\bigl( (U_\pi^{AAAA} \ox U_\sigma^{BBBB}) \xi^{\ox 4} \bigr)$.
  In Appendix~\ref{sec:diagram-orgy}, Lemma~\ref{lemma:diagram-bound} we show that 
  \[\begin{split}
    {\sum}_{\pi,\sigma \in S_4} \tr\bigl( (U_\pi^{AAAA} \ox U_\sigma^{BBBB}) \xi^{\ox 4} \bigr)      
                    &\leq 153(\tr(\xi^2))^2 + 126(\tr(\xi^2))(\tr(\xi_A^2)) + 126(\tr(\xi^2))(\tr(\xi_B^2)) \\
                    &\phantom{==}
                            + 9(\tr(\xi_A^2))^2 + 9(\tr(\xi_B^2))^2 + 30(\tr(\xi_A^2))(\tr(\xi_B^2))      \\
                    &\leq 153\bigl( \tr(\xi^2) + \tr(\xi_A^2) + \tr(\xi_B^2) \bigr)^2.
  \end{split}\]
  Plugging this into eq.~(\ref{eq:UU-4th-moment}), we find
  \begin{equation}
    \label{eq:UU-4th-moment-upper}
    \EE S^4 \leq \left( \frac{d_A d_B}{(d_A+1)(d_B+1)} \right)^3
                    153\bigl( \tr(\xi^2) + \tr(\xi_A^2) + \tr(\xi_B^2) \bigr)^2.
  \end{equation}
  Now we conclude as in the single-system case: by virtue of
  eqs.~(\ref{eq:UU-2nd-moment}) and~(\ref{eq:UU-4th-moment-upper}),
  \[\begin{split}
    \bigl\| ({\cal U}_A\ox{\cal U}_B)\xi \bigr\|_1
               &=    \EE |S| \\
               &\geq \sqrt{\frac{(\EE S^2)^3}{\EE S^4}} \\
               &\geq \frac{1}{\sqrt{153}} \sqrt{\tr(\xi^2) + \tr(\xi_A^2) + \tr(\xi_B^2)} \\
               &\geq \frac{1}{\sqrt{153}} \| \xi \|_2
                \geq \frac{1}{\sqrt{153 D}} \| \xi \|_1,
  \end{split}\]
  and we are done.
\end{proof}

\begin{remark}
  From the proof we see that, just as in the single-system case of
  Subsection~\ref{subsec:t-design}, it is enough for the local
  measurements to be $4$-designs.
\end{remark}

\begin{corollary}
  \label{cor:c-o-d-for-LO}
  The constants of domination, for locality-restricted measurements
  on a $d\times d$-system, are in the following relations:
  \begin{equation}
    \label{eq:chain1}
    \frac{1}{\sqrt{153} d} \leq \lambda\bigl( {\cal U}\ox{\cal U} \bigr)
                   \leq \lambda(\mathbb{LO})
                   \leq \lambda(\mathbb{LOCC})
                   \leq \lambda(\mathbb{SEP})
                   \leq \lambda(\mathbb{PPT})
                   \leq \frac{2}{d+1}.
  \end{equation}
  For separable measurements we have the even tighter bounds,
  \begin{equation}
    \label{eq:chain2}
    \frac{1}{d} \leq \lambda(\mathbb{SEP})
                \leq \lambda(\mathbb{PPT})
                \leq \frac{2}{d+1}.
  \end{equation}
\end{corollary}
\begin{proof}
  The first inequality in (\ref{eq:chain1}) is just Theorem~\ref{thm:per-aspera-ad-astram},
  the chain is by inclusion of the sets of POVMs, with the last bound
  following from the data hiding states $\alpha_d$ and $\sigma_d$, the
  (appropriately normalised) projections onto the (anti-)symmetric
  subspace of $\CC^d \ox \CC^d$ -- see~\cite{data-hiding1,data-hiding2}
  and~\cite{Matthews:Winter}.
  By Lemma~\ref{lemma:sep-bound} finally,
  $\lambda(\mathbb{SEP}) \geq \frac{1}{\sqrt{D}} = \frac{1}{d}$.
\end{proof}

\begin{remark}
  The first inequality~(\ref{eq:chain1}) in Corollary~\ref{cor:c-o-d-for-LO}
  proves a conjecture about the  optimal bias achievable with LOCC
  measurements (\cite[Conjecture 7]{Matthews:Winter}.
  Compare also with~\cite{data-hiding1}, where a bias of order $1/d^{2}$ was
  proven using a particular tomographically complete measurement, and it
  was suggested there that better POVMs might exist.
  
  This result shows that in a very strong sense the original data hiding
  states, the symmetric and anti-symmetric subspace projections, are
  essentially optimal: up to a constant factor they achieve the best
  available bias, which is $\Theta(1/d)$.
\end{remark}

\begin{remark}
  The $\ell^2$-bound in Theorem~\ref{thm:per-aspera-ad-astram} has another
  notable consequence for data hiding: observing that for orthogonal states
  $\rho$ and $\sigma$, 
  \[
    \| \rho-\sigma \|_2 =    \sqrt{\tr(\rho^2)+\tr(\sigma^2)}
                        \geq \max \left\{ \| \rho \|_2, \| \sigma \|_2 \right\},
  \]
  we conclude that data hiding states \emph{have} to be highly mixed.
  If one of them has rank bounded by $r$, say, Theorem~\ref{thm:per-aspera-ad-astram}
  places a lower bound of $1/13r$ on the bias achievable by LOCC measurements.
  
  Indeed, all known constructions of data hiding states endow them with
  considerable entropy (comparable to or larger than the size of the
  ``shares''), see~\cite{data-hiding1,data-hiding2,rand}. Our bound
  tells us that this has to be so to guarantee security of the scheme.
  We intend to return to this issue on a separate occasion.
\end{remark}

\section{Certainty relations}\label{sec:certainty}
The results on $\lambda({\cal M})$ for the isotropic POVM, tensor
products of isotropic POVMs, and $2$-designs have nice interpretations
as ``certainty relations'' in the sense of Sanchez-Ruiz~\cite{Sanchez-Ruiz:certainty}.
Namely, for a complete set of $d+1$ mutually unbiased bases in $\CC^d$ with associated
basis measurements ${\cal B}_k$, he
shows that for any pure state $\varphi = \proj{\varphi}$,
\begin{equation}\label{eq:ruiz}
  (d+1)\log\frac{d+1}{2} \leq \sum_{k=0}^d S_2\bigl( {\cal B}_k(\varphi) \bigr) \leq (d+1)\log d - \log(d-1),
\end{equation}
where $S_2({\cal B}) = - \log \sum_{x} |\inp{x}{\varphi}|^4$ is the R{\'e}nyi entropy of order 2
for the orthonormal basis ${\cal B} = \{\ket{1},\ldots\ket{d}\}$.
The right hand side of eq.~(\ref{eq:ruiz}) is referred to as a certainty relation, and 
intuitively states that for the chosen measurements
there exists no pure state that will lead to maximum entropy for all measurements simultaneously.
It quantifies the fact (quite natural, after a moment of thought)
that not all the tomographic data from measuring those bases 
is equally informative in the sense of Shannon. 
The certainty relation of~\cite{Sanchez-Ruiz:certainty} also holds for the Shannon entropy.
Let ${\cal M}$ be the measurement formed by
measuring in one of the $d+1$ bases at random. 
Using the concavity of the log, the certainty relation can then
be rewritten as
\[
  \log\bigl( d(d+1) \bigr) - S_2\bigl( {\cal M}(\varphi) \bigr) \geq \frac{1}{d+1}\log(d-1). 
\]

From our results in the previous section, we can infer similar certainty relations. 
First of all, from Theorem~\ref{thm:2-design} 
we get the following more general but weaker bound
for any proper $2$-design POVM with $n$ outcomes:
\[\begin{split}
  \log n - S_2\bigl( {\cal M}(\varphi) \bigr)
                                  &\geq \log n - S\bigl( {\cal M}(\varphi) \bigr)          \\
                                  &=    D\bigl( {\cal M}(\varphi) \| {\cal M}(\1/d) \bigr) \\
                                  &\geq \frac{1}{2\ln 2} \| {\cal M}(\varphi-\1/d) \|_1^2  \\
                                  &\geq \frac{1}{4\ln 2} \frac{d-1}{d(d+1)^2}
                                   \geq \frac{1}{6\ln 2} \frac{1}{(d+1)^2},
\end{split}\]
where the second inequality follows from the Pinsker inequality $D(\rho\|\sigma) \geq \frac{1}{2\ln 2}\|\rho-\sigma\|_1^2$.

For uni- and bipartite $4$-designs, in particular the isotropic POVMs,
we get considerably better bounds, due to the appearance of the Hilbert-Schmidt
norm. Consider any ensemble of quantum states, $\rho = \sum_x p_x \rho_x$.
For the Shannon mutual information between the preparation variable
$X$ (distributed according to $p_x$) and the measurement outcome given by ${\cal U}$,
\begin{equation}\label{eq:Iacc_LB}\begin{split}
  I(X:{\cal U}) &=    \sum_x p_x D\bigl( {\cal U}(\rho_x) \| {\cal U}(\rho) \bigr)                     \\
                &\geq \sum_x p_x \frac{1}{2\ln 2} \bigl\| {\cal U}(\rho_x) - {\cal U}(\rho) \bigr\|_1^2 \\
                &\geq \sum_x p_x \frac{1}{18\ln 2} \bigl\| {\cal U}(\rho_x) - {\cal U}(\rho) \bigr\|_2^2 \\
                &=    \frac{1}{18\ln 2}\left( \sum_x p_x \tr(\rho_x^2) - \tr(\rho^2) \right)
                 =    \frac{1}{18\ln 2}\left( S_L(\rho) - \sum_x p_x S_L(\rho_x) \right).
\end{split}\end{equation}
In other words, we get a lower bound on the accessible information of the ensemble
in terms of so-called ``linear entropies'' $S_L(\rho) = 1-\tr(\rho^2)$. In the above
derivation we have used the well-known relation between mutual information and
relative entropy, the Pinsker inequality
and eq.~(\ref{eq:AE}).

A particularly interesting case is that of a pure state ensemble $\rho_x = \proj{\varphi_x}$:
all the $S_L(\rho_x)$ are zero, so we get a positive lower bound for the accessible information
\[
  I_{\text{acc}}\bigl( \{p_x,\varphi_x\} \bigr) \geq I(X:{\cal U}) \geq \frac{1}{18\ln 2} \bigl( 1-\tr(\rho^2) \bigr),
\]
which is a small but positive constant, depending only on $\rho$.
It turns out that the best possible lower bound on the accessible information
in terms solely of $\rho$ is known: it is the so-called \emph{subentropy}
$Q(\rho)$ of Jozsa, Robb and Wootters~\cite{JRW-subentropy}, attained on a 
particular ensemble decomposition of $\rho$, named after Ebenezer Scrooge.
Incidentally, for this ensemble \emph{all} complete (i.e., rank-$1$) POVMs
have the same information gain. It is largest on the maximally mixed state,
and bounded by $\frac{1-\gamma}{\ln 2} \approx .6099$, where $\gamma$ is
Euler's constant~\cite{JRW-subentropy}.

For bipartite systems we furthermore obtain a lower bound
for $I^{\text{LOCC}}_{\text{acc}}(\cdot)$, that is the accessible
information when we are restricted to performing LOCC measurements. This bound is obtained
by using Theorem~\ref{thm:per-aspera-ad-astram} to lower bound $I(X:{\cal U}_A\ox{\cal U}_B)$
 -- the mutual information when the locally unitarily invariant continuous POVM is used.
This quantity is studied as a lower bound on the locally accessible information
in~\cite{SenDe-Sen-Lewenstein} (where it is denoted $\Lambda_L\L( \{p_x,\varphi_x\}\R)$).
Unlike the subentropy, this quantity depends on the ensemble (rather than the
ensemble average alone) even when it is a pure state ensemble.
However, in ~\cite{SenDe-Sen-Lewenstein} it is interpreted differently 
as the average of the mutual information over all complete product basis measurements.
Since some measurements of this form cannot be performed by LOCC, the
authors (unnecessarily) restrict their claim that it is a lower bound on the locally
accessible information to bipartite systems of $2 \times n$ dimensions (where
it is known that any complete product basis measurement can be performed by LOCC).
This is unnecessary because, as described in Section \ref{sec:local-POVM}, $I(X:{\cal U}_A\ox{\cal U}_B)$
is also the mutual information yielded by the protocol where Alice and Bob independently measure
according to the unitarily invariant continuous POVM and share their results
(which is clearly accomplished by LOCC). As noted in~\cite{SenDe-Sen-Lewenstein},
this bound is saturated by Scrooge ensembles.

No general closed form is known for $I(X:{\cal U}_A\ox{\cal U}_B)$
(although some special cases are derived in~\cite{SenDe-Sen-Lewenstein}) so it is useful
to note that by using the same derivation as in (\ref{eq:Iacc_LB}), but invoking Theorem~\ref{thm:per-aspera-ad-astram}, we get
that for an arbitrary ensemble on a bipartite system,
\begin{equation}
 I^{\text{LOCC}}_{\text{acc}}\bigl( \{p_x,\rho_x\} \bigr)
            \geq I(X:{\cal U}_A\ox{\cal U}_B)
            \geq \frac{1}{306\ln 2} \left( S_L(\rho) - \sum_x p_x
S_L(\rho_x) \right).
\end{equation}

It is worth noting that in the case of an ensemble of pure states this lower 
bound, unlike $I(X:{\cal U}_A\ox{\cal U}_B)$, depends only on the ensemble average. 
Hence we get a lower bound of
\[
  Q^{\text{LOCC}}(\rho) 
     :=   \inf_{\rho = \sum_x p_x \varphi_x} I^{\text{LOCC}}_{\text{acc}}\bigl( \{p_x,\varphi_x\} \bigr)
     \geq \frac{1}{306\ln 2} \left( 1- \tr(\rho^2) \right)
\]
on the LOCC-subentropy of $\rho$.

\section{Conclusion}
\label{sec:conclusion}
We have introduced a formalism of norms on states/density operators linked
to their (pairwise) distinguishability by a given, restricted, class of
measurements. This allows us to study the relation between these norms in
convex geometric terms.
We went on to investigate the constants of domination for the resulting norms
with respect to the well-known trace norm: for a single measurement we looked at the isotropic
POVM, $4$- and $2$-designs. Furthermore, we considered several classes of
locally restricted measurements, such as LOCC or PPT POVMs. The results
here have strong connection to data hiding: indeed, we proved that
up to a constant factor the hiding states of~\cite{data-hiding1}
achieve already the best possible bias.
We leave many questions open, such as the eventual determination of the
locally accessible information and better bounds on the constants of domination.
More importantly, one ought to be able to obtain more information on the geometry
of the convex bodies $\mathbb{M}$ and the unit balls of $\|\cdot\|_{\mathbb{M}}$ -- here
we only compared them with the trace and the Hilbert-Schmidt norms, but it would
be interesting to get more insight into their geometric shape.
It is an intriguing open question regarding single measurements where to place
$3$-design POVMs relative to $2$- and $4$-designs.


\acknowledgments
AW thanks the members of the Pavia Quantum Information group for an
enjoyable afternoon in October 2007, where he had occasion to discuss
some of the questions of the present paper, when they were still in
a nascent state. In particular the feedback of G. M. D'Ariano,
G. Chiribella and M. F. Sacchi, and their suggestions regarding the use of
symmetry, are gratefully acknowledged.
Ashley Montanaro provided the pointer to the paper by Ambainis and Emerson,
and provided the example mentioned in appendix~\ref{sec:l1}.
WM would like to thank Dan Shepherd for a useful discussion about groups and diagrams.

WM was supported by the U.K.~EPSRC.
SW was supported by NSF grant number PHY-04056720.
AW was supported by the U.K. EPSRC through the ``QIP IRC'' and an
Advanced Fellowship, by a Royal Society Wolfson Merit Award
and by the European Commission through IP ``QAP''. The Centre for
Quantum Technologies is funded by the Singapore Ministry of Education
and the National Research Foundation as part of the Research Centres
of Excellence programme.

\appendix

\section{An $\ell_1$-inequality for probability vectors and density operators}
\label{sec:l1}

\begin{proposition}
  \label{prop:l1-ineq}
  For probability vectors $\vec{p}$, $\vec{q}$ in $\RR^n$ (i.e.~$p_i \geq 0$
  and $\sum_{i=1}^n p_i=1$, and likewise for $q_i$),
  \begin{equation}
    \label{eq:conj}
    \| \vec{p} - \vec{q} \|_1 \geq 1 - n\, \vec{p}\cdot\vec{q},
  \end{equation}
  where on the left is the statistical distance between the distributions,
  namely the $\ell_1$-norm of their difference, and on the right we
  have the usual Euclidean inner product of vectors.
\end{proposition}

\begin{corollary}[Quantum case]
  Ineq.~(\ref{eq:conj}) has a
  straightforward quantum generalisation: for any two density operators
  $\rho$ and $\sigma$ on an $n$-dimensional Hilbert space,
  \begin{equation}
    \label{eq:conj-q}
    \| \rho - \sigma \|_1 \geq 1 - n\, \tr(\rho\sigma),
  \end{equation}
  where now on the left is the trace norm, and on the right
  is the Hilbert-Schmidt inner product on operator space.
\end{corollary}
This actually follows from the classical case, as follows: $\rho$ is
diagonalised in some basis, with a probability vector $\vec{p}$ along the
diagonal. Denote the dephasing operation in this basis by ${\cal E}$ -- it is
a CPTP map with ${\cal E}(\rho) = \rho$. Denoting $\sigma' = {\cal E}(\sigma)$,
which is now diagonalised in the same basis, with a probability vector
$\vec{q}$ along the diagonal, we now have
\[
  \frac{1}{2} \| \rho - \sigma \|_1 \geq \frac{1}{2} \| \rho - \sigma' \|_1
  \text{ and }
  \tr(\rho\sigma) = \tr(\rho\sigma'),
\]
so all we need to prove is
\[
  \frac{1}{2} \| \rho - \sigma' \|_1 \geq 1 - n\, \tr(\rho\sigma').
\]
But because of
\[
  \frac{1}{2} \| \rho - \sigma' \|_1 = \frac{1}{2} \| \vec{p} - \vec{q} \|_1
  \text{ and }
  \tr(\rho\sigma') = \vec{p}\cdot\vec{q}
\]
this is precisely~(\ref{eq:conj}).
\qed

\bigskip
\begin{proof}[of Proposition \ref{prop:l1-ineq}]
We use the well-known relation between trace distance and
fidelity~\cite{Fuchs:vandeGraaf}:
\[
  \frac{1}{2} \| \vec{p} - \vec{q} \|_1 \geq 1 - \sum_i \sqrt{p_i q_i},
\]
hence we are done once we show
\[
  2\left( 1 - \sum_i \sqrt{p_i q_i} \right) \geq 1 - n \sum_i p_i q_i,
\]
which -- introducing the shorthand $t_i = \sqrt{p_i q_i}$ -- is equivalent to
\[
  \sum_i t_i  \leq \frac{1}{2} + \frac{1}{2} n \sum_i t_i^2.
\]
Now, for fixed $s = \sum_i t_i \leq 1$, the right hand side here is
minimal for $t_1 = \ldots = t_n = \frac{s}{n}$, in which case it reduces
to $\frac{1}{2} + \frac{1}{2} s^2$, which is indeed always $\geq s$. 
\end{proof}

\begin{remark}
  Ineq.~(\ref{eq:conj}) becomes false when introducing
  a factor $c<1$ on the left hand side. for sufficiently large $n$.
  Ashley Montanaro [personal communication] pointed out to us
  the following class of examples:
  
  Consider $\vec{p} = \left(x,0,\frac{1-x}{n-2},\ldots,\frac{1-x}{n-2}\right)$
  and $\vec{q} = \left(0,x,\frac{1-x}{n-2},\ldots,\frac{1-x}{n-2}\right)$,
  which have $c\| \vec{p}-\vec{q} \|_1 = 2cx$, whereas
  $1-n\vec{p}\cdot\vec{q} = 1-\frac{n}{n-2}(1-x)^2 \sim 2x+x^2$ for large $n$.
\end{remark}

\section{An integral over the unit sphere} 
\label{sec:integral}

\begin{lemma}
  \label{lemma:integral}
  Let $P$ and $Q$ be mutually orthogonal projectors of rank $a$ and $b$, respectively,
  in $\CC^d$. Then, for the uniform distribution on the unit vectors 
  $\ket{\psi} = \sum_{j=1}^d \psi_j \ket{j} \in \CC^d$,
  \[\begin{split}
    \EE \left| \frac{1}{2a}\tr(\psi P) - \frac{1}{2b}\tr(\psi Q) \right| 
         &= d \int {\rm d}\psi \left| \frac{1}{2a}\sum_{j=1}^a |\psi_j|^2 
                                     - \frac{1}{2b}\sum_{j=a+1}^{a+b} |\psi_j|^2 \right| \\
         &= 1 - \frac{1}{a+b}\sum_{k=0,\ldots a-1 \atop \ell=0,\ldots b-1}
                                           p^k (1-p)^\ell {k+\ell \choose k},
  \end{split}\]
  where $p=a/(a+b)$.
\end{lemma}
\begin{proof}
  Introduce a random Gaussian vector
  $\ket{\varphi} \sim {\cal N}_{\CC^d}(0,1)$~\cite{rsp},
  i.e.~$\ket{\varphi} = \frac{1}{\sqrt{2d}} \sum_{j=1}^d (\alpha_j + i \beta_j)\ket{j}$
  with independent Gaussian distributed real and imaginary parts
  $\alpha_j, \beta_j \sim {\cal N}(0,1)$ of zero mean and unit variance.
  In particular, $\EE \bra{\varphi} \varphi \rangle = 1$.
  
  Now, using this and the unitary invariance of the distribution 
  of $\ket{\varphi}$, we see
  \[\begin{split}
    \EE \left| \frac{1}{2a}\tr(\psi P) - \frac{1}{2b}\tr(\psi Q) \right| 
         &= \EE_\varphi \left( \bra{\varphi} \varphi \rangle 
                               \EE_\psi \left| \frac{1}{2a}\tr(\psi P) - \frac{1}{2b}\tr(\psi Q) \right|
                        \right)                                                             \\
         &= \EE_\varphi \left| \frac{1}{2a}\tr(\varphi P) - \frac{1}{2b}\tr(\varphi Q) \right|  \\
         &= \frac{1}{2} \EE_{\alpha_j,\beta_j \sim {\cal N}(0,1)}
                              \left| \frac{1}{2a}\sum_{j=1}^a (\alpha_j^2 + \beta_j^2)
                                     - \frac{1}{2b}\sum_{j=a+1}^{a+b} (\alpha_j^2 + \beta_j^2) \right| \\
         &= \frac{1}{2} \EE_{X,Y} \left| \frac{1}{2a}X - \frac{1}{2b}Y \right|.
  \end{split}\]
  The sums of squares of Gaussian components occurring here are well-studied,
  and known under the name of $\chi^2$-distributions:
  \[
    \sum_{j=1}^a (\alpha_j^2 + \beta_j^2) =: X \sim \chi^2_{2a},\quad
    \sum_{j=a+1}^{a+b} (\alpha_j^2 + \beta_j^2) =: Y \sim \chi^2_{2b},
  \]
  their probability density being given by
  \[
    \Pr\{ X\in [x;x+{\rm d}x] \} = \frac{1}{2 (a-1)!} (x/2)^{a-1} e^{-x/2} {\rm d}x,\quad
    \Pr\{ Y\in [y;y+{\rm d}y] \} = \frac{1}{2 (b-1)!} (y/2)^{b-1} e^{-y/2} {\rm d}y.
  \]
  
  This allows us to evaluate the latter expectation as follows, denoting the
  indicator function of a set $\{\ldots\}$ as $\mathbf{1}\{\ldots\}$:
  \[\begin{split}
    \frac{1}{2} \EE_{X,Y} \left| \frac{1}{2a}X - \frac{1}{2b}Y \right|
          &= \frac{1}{2} \EE_{X,Y}
                           \left( \int {\rm d}r \,\mathbf{1}\{X/2a \leq r \leq Y/2b\}
                                  + \int {\rm d}r \,\mathbf{1}\{Y/2b \leq r \leq X/2a\} \right) \\
          &= \frac{1}{2} \int_0^\infty {\rm d}r
                           \bigl( \EE \,\mathbf{1}\{X \leq 2ar, Y \geq 2br\} 
                                  + \EE \,\mathbf{1}\{X \geq 2ar, Y \leq 2b\} \bigr)    \\
          &= \frac{1}{2} \int_0^\infty {\rm d}r
                           \bigl( \Pr\{X\leq 2ar\} \Pr\{Y\geq 2br\}
                                  +  \Pr\{X\geq 2ar\} \Pr\{Y\leq 2br\} \bigr) \\
          &= \frac{1}{2} \int_0^\infty {\rm d}r \Pr\{X\geq 2ar\}
             + \frac{1}{2} \int_0^\infty {\rm d}r \Pr\{Y\geq 2br\}            \\
          &\phantom{=:}
             - \int_0^\infty {\rm d}r \Pr\{X\geq 2ar\}\Pr\{Y\geq 2br\}.
  \end{split}\]
  Using the $\chi^2$ densities, the probabilities under the integrals are easily
  evaluated:
  \[
    \Pr\{X\geq 2ar\} = e^{-ar} \sum_{k=0}^{a-1}    \frac{(ar)^k}{k!},\quad
    \Pr\{Y\geq 2br\} = e^{-br} \sum_{\ell=0}^{b-1} \frac{(br)^\ell}{\ell!}.
  \]
  This finally gives
  \[\begin{split}
    \EE \left| \frac{1}{2a}\tr(\psi P) - \frac{1}{2b}\tr(\psi Q) \right| 
          &= \frac{1}{2} \EE_{X,Y} \left| \frac{1}{2a}X - \frac{1}{2b}Y \right| \\
          &= \frac{1}{2} + \frac{1}{2}
             - \int_0^\infty {\rm d}r \,e^{-r(a+b)} \!\!\!\! 
                   \sum_{k=0,\ldots a-1 \atop \ell=0,\ldots b-1} \frac{(ar)^k (br)^\ell}{k!\ell!} \\
          &= 1 - \frac{1}{a+b}\sum_{k=0,\ldots a-1 \atop \ell=0,\ldots b-1}
                    \left(\frac{a}{a+b}\right)^k \left(\frac{b}{a+b}\right)^\ell {k+\ell \choose k},
  \end{split}\]
  where we have used the integral for the Gamma function.
\end{proof}

We will also need the following small lemma
\begin{lemma}\label{lemma:binomial}
       Let $S_k$ denote $\sum_{l=0}^{k} 2^{-(k+l)} \binom{k+l}{l}$. We claim
that for integers $k \geq 0$, $S_k = 1$.
\end{lemma}
\begin{proof}
       Using the well known 'addition formula' $\binom{n}{m} =
\binom{n-1}{m}+\binom{n-1}{m-1}$
       \begin{align}
               S_{k+1} &= \sum_{l=0}^{k+1} 2^{-(1+k+l)} \binom{k+l}{l} +
\sum_{l=0}^{k+1} 2^{-(1+k+l)} \binom{k+l}{l-1} \\
               &= \frac{1}{2} S_{k} + 2^{-(2k + 2)} \binom{2k+1}{k+1} +
\sum_{l=0}^{k} 2^{-(2+k+l)} \binom{k+l+1}{l} \\
       &= \frac{1}{2} S_{k} + 2^{-(2k + 2)} \binom{2k+1}{k+1} +
\frac{1}{2}S_{k+1} - \frac{1}{2} 2^{-(2k + 2)}\binom{2k+2}{k+1}
       \end{align}
       so
       \[
               S_{k+1} = S_{k} + 2^{-(2k + 2)}\L( 2 \binom{2k+1}{k+1} -
\binom{2k+2}{k+1} \R) =  S_{k}
       \]
where the final equality is due to the addition formula and the
symmetry $\binom{2k+1}{k+1} = \binom{2k+1}{k-1}$.
       To complete the proof we note that $S_{0} = 1$.
\end{proof}


\section{Upper bounds on certain traces}
\label{sec:diagram-orgy}

\begin{lemma}
  \label{lemma:diagram-bound}
  Let $\xi$ be a traceless Hermitian operator on a bipartite 
  Hilbert space $\mathcal{H}_A \ox \mathcal{H}_B$. Let $P^{(4)}_{{\rm sym}\,A}$ and
  $P^{(4)}_{{\rm sym}\,B}$ denote the projector onto the completely symmetric
  subspace of $\mathcal{H}_A^{\ox 4}$ and $\mathcal{H}_B^{\ox 4}$, respectively.

  Then, with the shorthands $t := \tr (\xi^2)$, $a := \tr (\xi_A^2)$ and $b := \tr (\xi_B^2)$,
  where $\xi_A = \tr_B (\xi)$ and $\xi_B = \tr_A (\xi)$,
  \begin{equation}\label{4thMomentUB}
    \tr \left(\bigl(P^{(4)}_{{\rm sym}\,A}\ox P^{(4)}_{{\rm sym}\,B}\bigr) \xi^{\ox 4}\right)  
            \leq \frac{1}{4!^2}\bigl(153 t^2 + 126 ta + 126 tb + 9 a^2 + 9 b^2 + 30ab \bigr).
  \end{equation}
\end{lemma}

The proof is conceptually simple but a little long. We write the projection 
operators as averages over the unitary operators which permute 
the four subsystems. Defining, for permutations $\pi \in S_4$, the representation
\[
	U^{A}_{\pi} := \sum_{{\bf j} \in \{1,\ldots,d\}^m } \Ox_{i=1}^{4} \ket{j_i}^{A}_{\pi(i)}\bra{j_i}^{A}_{i}
\]
where $\{\ket{j}^{A}_i\}_{1\leq j \leq d}$ is an orthonormal basis for the $i$-th copy of $\mathcal{H}_A$ in $\mathcal{H}_A^{\ox 4}$,
and defining $U^{B}_{\pi}$ similarly:

\[
	\tr \left(\bigl(P^{(4)}_{{\rm sym}\,A}\ox P^{(4)}_{{\rm sym}\,B}\bigr) \xi^{\ox 4} \right)
	= \frac{1}{24!^2}\displaystyle\sum_{\p\in S_4, \s\in S_4} \tr \left(U^{A}_{\p}\ox U^{B}_{\s} \xi^{\ox 4}\right)
\]

Clearly $(\p, \s) \to U^{A}_{\p}\ox U^{B}_{\s}$ is a representation of $S_4 \times S_4$. $S_4 \times S_4$ has
a subgroup consisting of all the elements of the form $(g, g)$, which we'll denote by $R$.

If $(\p', \s') = r^{-1} (\p, \s) r$ for some $r \in R$, 
we write $(\p', \s') \stackrel{R}{\sim} (\p, \s)$ and note
that the corresponding terms are equal since

\[
	\tr \left(U^{A}_{\p'}\ox U^{B}_{\s'} \xi^{\ox 4} \right)
	   = \tr \left((U^{A}_{\p}\ox U^{B}_{\s})
	                  (U^{A}_{g}\ox U^{B}_{g})\xi^{\ox 4}(U^{A}_{g^{-1}}\ox U^{B}_{g^{-1}})\right)
	   = \tr \left(U^{A}_{\p}\ox U^{B}_{\s} \xi^{\ox 4}\right).
\]

Essentially, conjugation by an element of $R$ corresponds to a permutation 
of the identical $\xi$ operators, and therefore leaves the term unchanged.

The set of all $24!^2$ terms is partitioned by the equivalence relation $\stackrel{R}{\sim}$ with
the terms in each subset all equal to each other. We shall refer to these subsets as the
$R$-conjugacy classes of $S_4 \times S_4$. Clearly, the $R$-conjugacy classes form a finer
partition of $S_4 \times S_4$ than the normal conjugacy classes.

By demonstrating an appropriate upper-bound for
the terms in each $R$-conjugacy class, and calculating the size of each class,
we will prove the upper bound (\ref{4thMomentUB}).

\medskip\noindent		
\emph{Tensor Diagrams.}
Let establish an orthonormal basis $\{ \ket{i}_A \}$ ($\{ \ket{i}_B \}$) for $\mathcal{H}_A$ ($\mathcal{H}_B$).
In this basis, we can write $\xi$ in component form thus $\xi_{i,j}^{k,l} = \bra{k}_A \ox \bra{l}_B \xi \ket{i}_A \ox \ket{j}_B$

We would like to demonstrate upper bounds for terms of the form
\begin{equation}\label{eq:index-perm}
	\tr \left(U^{A}_{\p}\ox U^{B}_{\s} \xi^{\ox 4}\right) = \xi_{a1,b1}^{a\p(1),b\s(1)} 
	                                            \xi_{a2,b2}^{a\p(2),b\s(2)} 
	                                            \xi_{a3,b3}^{a\p(3),b\s(3)} 
	                                            \xi_{a4,b4}^{a\p(4),b\s(4)},
\end{equation}
where the $ai$ and $bi$ ($i \in \{1,2,3,4\}$) are dummy variables to by contracted over 
according to the Einstein summation convention.
Using indices in our calculations would be rather messy and confusing. Instead we use the
ingenious tensor diagrams of Penrose~\cite{penrose}:

We denote our bipartite Hermitian operator $\xi$ by 
\inlinediagm{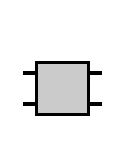}{0.3in}. 
The ``terminals'' of this diagram correspond to 
indices like so
\[
	\xi_{i,j}^{k,l} = \inlinediagm{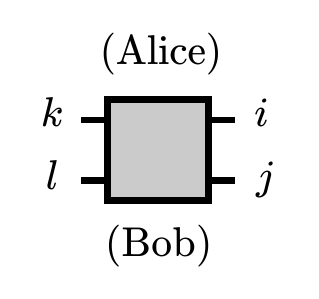}{0.6in}.
\]
Joining the terminals with ``wires'' denotes contraction 
of the corresponding indices
\[
	\xi_{r,j}^{k,l} \xi_{p,q}^{r,m} = \inlinediagm{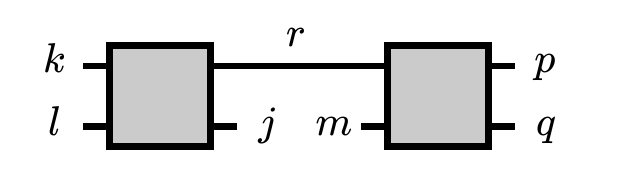}{1.3in}
\]
\begin{align*}
	\begin{array}{cc}
		\xi_A := \tr_B (\xi) = \inlinediagm{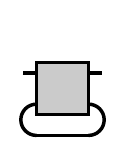}{0.3in},	& \xi_B := \tr_A (\xi) = \inlinediagm{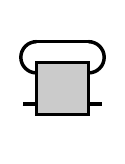}{0.3in},\\
		\tr (\xi) = \inlinediagm{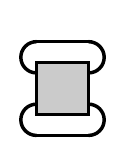}{0.3in} = 0,		& \tr (\xi^2) = \inlinediagm{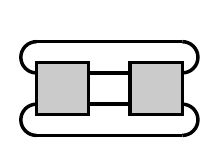}{0.45in} = t,\\
		\tr (\xi_A^2) = \inlinediagm{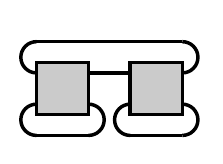}{0.45in} = a,	& \tr (\xi_B^2) = \inlinediagm{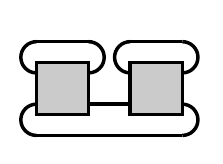}{0.45in} = b.
	\end{array}
\end{align*}
	
In an effort to keep the diagrams tidy and compact, we sometimes 
use a pair of vertical grey lines, one with wires entering from the 
right and the other with a matching set of wires entering from the 
left. A diagram with this feature is to be read as equivalent to the 
diagram one obtains by identifying the grey lines in parallel to 
join the matching wires. It should not be confused with the bars 
drawn \emph{across} wires (by Penrose and others) to denote (anti-)symmetrization.

Here is an example showing how a diagram corresponds to a particular term of the form (\ref{eq:index-perm}):
\[
\inlinediagm{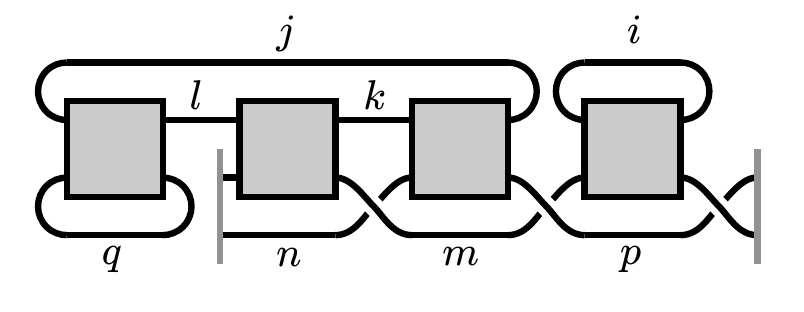}{1.7in} = \xi_{l,q}^{j,q} \xi_{k,m}^{l,p} \xi_{j,p}^{k,n} \xi_{i,n}^{i,m}
\]

\begin{figure}
  \includegraphics[scale=0.7]{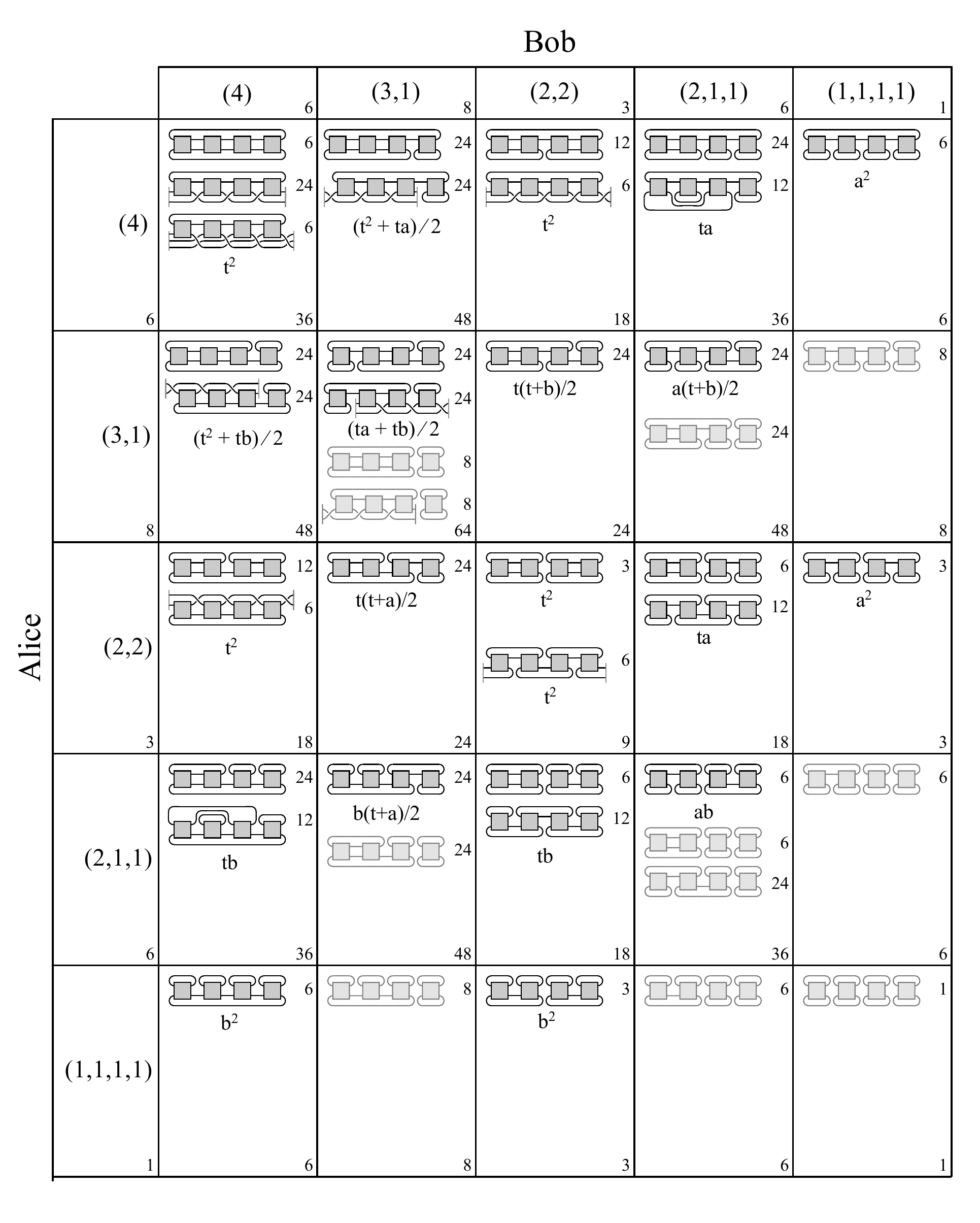}
  \caption{Sizes and upper-bounding expressions of the R-conjugacy classes. The faded diagrams are identically zero (because they contain a factor of $\tr(\xi)$).}
  \label{fig:one}
\end{figure}

In Fig.~\ref{fig:one} we provide a table with a diagram representative
of each of the $R$-conjugacy classes organised
by the conjugacy class of $S_4 \times S_4$ which contains it.

The size of each $R$-conjugacy class is written to the right of the corresponding diagram.
An upper bound is given and diagrams which are 
identically 0 (by virtue of having a factor of $\tr (\xi) = 0$) are drawn 
in a lighter shade of grey.

\medskip\noindent
\emph{Proofs of upper bounds.}
We give bounds for the terms shown in the upper-right triangle of 
Fig.~\ref{fig:one}.
Bounds for those terms below the diagonal follow from these 
by exchanging the roles of the parties. 
We will make repeated use of the Cauchy-Schwarz inequality 
for the Hilbert-Schmidt inner product,
\begin{lemma}
  $|\tr (A\hc B) |^2 \leq (\tr (A\hc A)) (\tr (B\hc B))$. \qed
\end{lemma}
Let $P$ denote a positive semidefinite hermitian operator. We have 
the inequality $\tr (P^2) \leq (\tr (P))^2$ (by the spectral 
decomposition of $P$ for example). From this fact and the 
Cauchy-Schwarz inequality it follows that
\begin{lemma}\label{lemma:psdIP}
  If $P$ and $Q$ are both positive semidefinite, then
  $\tr (PQ)  \leq (\tr (P))(\tr (Q))$.
  \qed
\end{lemma}
Third, since the partial transpose map is selfadjoint,
\begin{lemma}\label{lemma:pt}
  The quantities $t$, $a$ and $b$ are unchanged if we replace $\xi$ with $\xi\pt$.
  \qed
\end{lemma}

\bigskip
\begin{proof}[of Lemma~\ref{lemma:diagram-bound}]
We go through the types one by one.

\begin{list}{}{\itemindent=0cm}
  \item[{\bf (2,2):(2,2)}]
    $\inlinediag{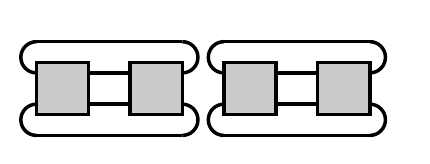} = (\tr (\xi^2))^2 = t^2$.
    To show that the same bound applies to $\inlinediag{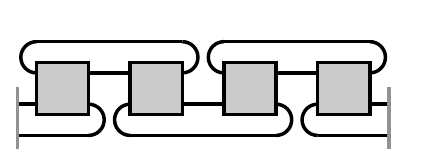}$,
    we note that it can be written as $\tr\bigl((\tr_A (Z))^2\bigr)$, where
	\[
		Z = (\xi\ox\1_C) (\1_A\ox \ketbra{\Phi}{\Phi}) (\xi\ox\1_C)
	\]
	and $\ket{\Phi} = \sum_{i=1}^{d}\ket{i}_B\ox\ket{i}_C$.
	Since $Z = \bigl( (\xi\ox\1_C)(\1_A\ox\ket{\Phi}) \bigr)
	             \bigl( (\xi\ox\1_C)(\1_A\ox\ket{\Phi}) \bigr)^{\dag}$,
	it is positive semidefinite and as such 
	$\tr\bigl((\tr_A (Z))^2\bigr) \leq (\tr (Z))^2$. The result follows by noting that $\tr (Z) = t$.

  \item[{\bf (2,2):(1,1,1,1)}]
	$\inlinediag{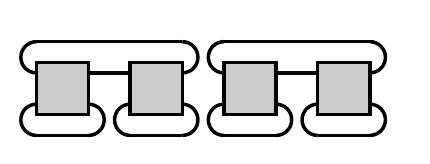} = (\tr (\xi_A^2))^2 = a^2$.
	
  \item[{\bf (2,1,1):(2,1,1)}]
	$\inlinediag{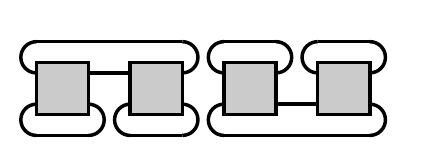} = ab$.

  \item[{\bf (4):(4)}]
	Noting that $\xi^2$ is positive semidefinite, and applying Lemma \ref{lemma:psdIP}, 
	we get $\inlinediag{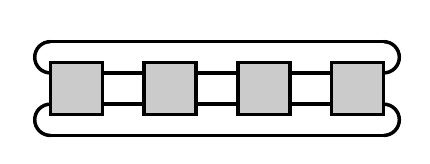} = \tr (\xi^4) \leq (\tr (\xi^2))^2 = t^2$.
	The partial-transpose of $\xi$, $\xi\pt$, has the diagrammatic representation
	$\inlinediagm{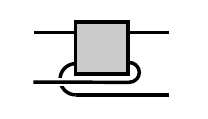}{0.5in}$ (we choose to take the transpose on Bob's system).
	Substituting, this for $\xi$ in $\inlinediag{figs/4-4A.pdf}$ results in the diagram
	$\inlinediag{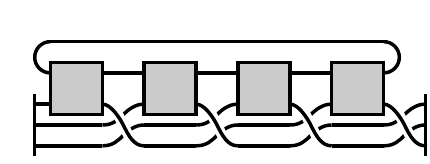} = (\tr (\xi\pt))^4$, so Lemma \ref{lemma:pt} shows 
	that the same bound applies here. The Cauchy-Schwarz inequality yields 
	$\inlinediag{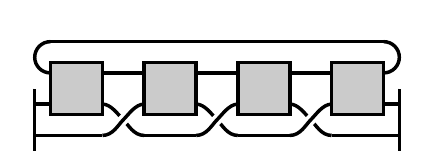} = \tr\bigl( (\xi\pt)^2 (\xi^2)\pt \bigr) 
	           \leq \sqrt{ (\tr (\xi\pt))^4 (\tr\bigl((\xi^2)\pt\bigr))^2 } 
	           =    \L(\inlinediag{figs/4-4A.pdf} \cdot \inlinediag{figs/4-4C.pdf}\R)^{1/2}$, 
	which can be seen to be $\leq t^2$ because of the previous two bounds.
	
  \item[{\bf (4):(2,1,1)}]
	$\inlinediag{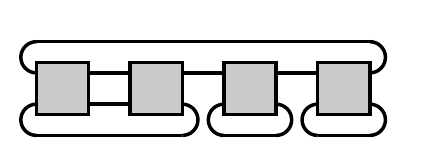} = \tr\bigl( (\tr_B (\xi^2)) \xi_A^2 \bigr) 
	                              \leq (\tr (\xi^2)) (\tr (\xi_A^2)) = ta$, by Lemma \ref{lemma:psdIP}.	%
	$\inlinediag{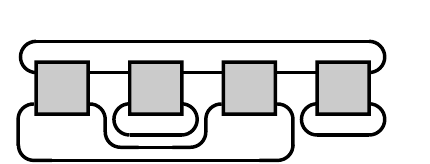} = \tr\bigl( \xi (\xi_A\ox\1_B) \xi (\xi_A\ox\1_B) \bigr)
	                              \leq \tr\bigl( \xi (\xi_A^2\ox\1_B) \xi \bigr)
	                              = \inlinediag{figs/4-211A.pdf}$,
	where we have used the Cauchy-Schwarz inequality.

  \item[{\bf (4):(3,1)}]
	$\inlinediag{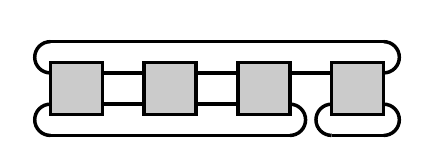}
	  \leq \L( \inlinediag{figs/4-4A.pdf} \cdot \inlinediag{figs/4-211A.pdf}\R)^{1/2}$. 
	Using the results for these two diagrams and the arithmetic-geometric 
	mean inequality we can bound this expression by $t(t+a)/2$, as was claimed.
	$\inlinediag{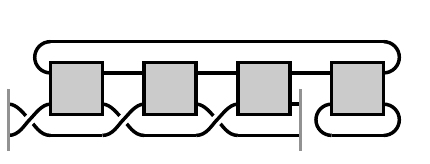}$ is given by substituting $\xi\pt$ into the 
	previous diagram, so by Lemma \ref{lemma:pt} the previous bound applies.

  \item[{\bf (4):(2,2)}]
	$\inlinediag{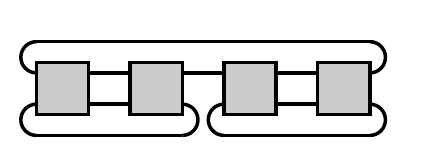} = \tr(\tr_B (\xi^2))^2 \leq t^2$.
	For the other diagram we use the Cauchy-Schwarz inequality:
	$\inlinediag{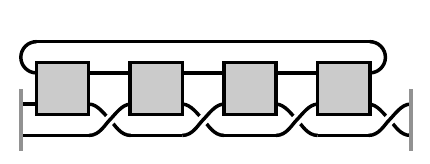} 
	 \leq \L(\L(\tr\L(\inlinediagm{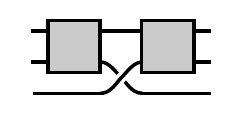}{0.5in}\cdot\inlinediagm{figs/XFX.pdf}{0.5in})\R)\L(
	         \tr(\inlinediagm{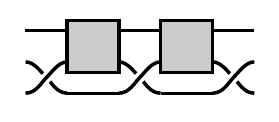}{0.5in}\cdot\inlinediagm{figs/FXFXF.pdf}{0.5in}\R)\R)
	      \R)^{1/2}=\inlinediag{figs/4-22A} \leq t^2$.
	
  \item[{\bf (2,2):(2,1,1)}]
	$\inlinediag{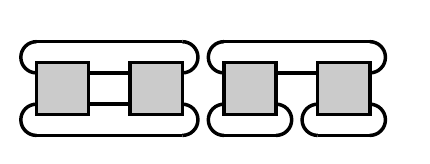} = ta$.
	For the other diagram in this class it is useful to define 
	$Y_B := \tr_A\bigl( \xi (\xi_A\ox\1_B) \bigr)$. 
	We define $Y_A$ similarly but with the roles of the parties reversed.
	\begin{align*}
		\tr (Y_B^2) &= \tr\Bigl( \left(\tr_A\bigl( \xi (\xi_A\ox\1_B) \bigr)\right) \cdot Y_B \Bigr)\\
		&= \tr\bigl( \xi ( \xi_A\ox\1_B ) (\1_A \ox Y_B) \bigr) \\
		&= \tr\bigl( \xi (\xi_A \ox Y_B) \bigr)\\
		&\leq \sqrt{ (\tr (\xi^2)) (\tr (\xi_A^2)) (\tr (Y_B^2) ) },
	\end{align*}
	and therefore
	\[
		\tr (Y_B^2) \leq (\tr (\xi^2))(\tr (\xi_A^2)) = ta.
	\]
	Similarly $\tr (Y_A^2) \leq tb$.
	Hence, $\inlinediag{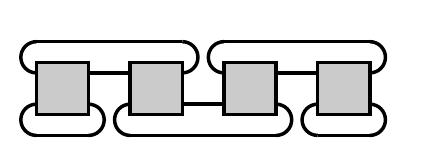} = \tr (Y_B^2) \leq ta$.

  \item[{\bf (4):(1,1,1,1)}]
	$\inlinediag{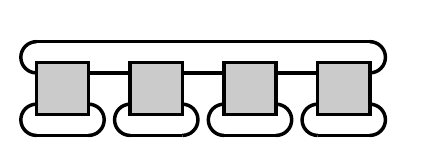} = \tr (\xi_A^4) \leq (\tr (\xi_A^2))^2 = a^2$.
	
  \item[{\bf (3,1):(3,1)}]
	$\inlinediag{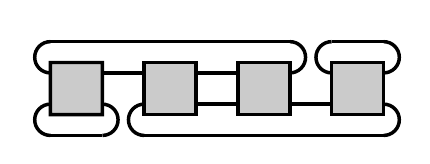} = \tr\bigl( (\xi_A \ox \1_B) \xi^2 (\1_A\ox \xi_B) \bigr) 
	                              = \tr\bigl( \xi^2 (\xi_A\ox \xi_B) \bigr)$.
	Using the Cauchy-Schwarz inequality
	we upper bound this by $\sqrt{ (\tr (\xi^4)) (\tr (\xi_A^2)) (\tr (\xi_B^2)) }$,
	which in turn is bounded by $(\tr (\xi^2)) \sqrt{ (\tr (\xi_A^2)) (\tr (\xi_B^2)) } \leq (ta + tb)/2$,
	using  arithmetic-geometric mean inequality at the end.
	$\inlinediag{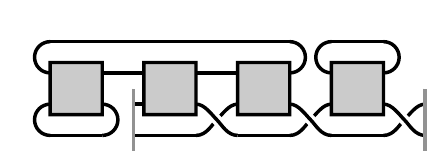}$ is given by substituting $\xi\pt$ into the 
	previous diagram, so by Lemma \ref{lemma:pt} the same bound applies.
	
  \item[{\bf (3,1):(2,2)}]
	$\inlinediag{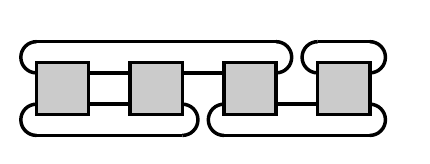} = \tr\bigl( (\tr_B (\xi^2) ) Y_A \bigr) 
	          \leq \sqrt{ (\tr\left(\tr_B (\xi^2))\right)^2 (\tr (Y_A^2)) } \leq t(t+b)/2$.
	
  \item[{\bf (3,1):(2,1,1)}]
	$\inlinediag{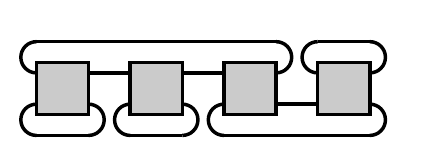} = \tr( \xi_A^2 Y_A ) \leq \sqrt{(\tr(\xi_A^4))(\tr (Y_A^2))} \leq a(t+b)/2$.
\end{list}

Now, collecting terms according to the multiplicities found in the table 
of Fig.~\ref{fig:one}, we conclude the proof.
\end{proof}

\begin{remark}
  Note that for every pair of conjugacy classes of permutations, all
  the types falling into the corresponding box in Fig.~\ref{fig:one}
  share the same upper bound.
\end{remark}

\end{document}